\documentclass[aps,prc,twocolumn,superscriptaddress]{revtex4}

\usepackage{graphicx}
\usepackage{draftcopy}
\usepackage{afterpage}

\newcommand{\Eq}[1]   {Eq.~(\ref{#1})}

\newcommand{\Fi}[1]   {Fig.~\ref{#1}}
\newcommand{\Fis}[2]  {Figs.~\ref{#1} and~\ref{#2}}
\newcommand{\Ta}[1]   {Table~\ref{#1}}

\newcommand{\agev}    {\mbox{$A$~GeV}}               
\newcommand{\gevc}    {\mbox{GeV$/c$}}
\newcommand{\gevcc}   {\mbox{GeV$/c^2$}}

\newcommand{\mevcc}   {\mbox{MeV$/c^2$}}
\newcommand{\rb}[1]   {\mbox{\textrm{\scriptsize #1}}}
\newcommand{\rbt}[1]  {\mbox{\textrm{\tiny #1}}}
\newcommand{\sqrts}   {\ensuremath{\sqrt{s_{_{\rbt{NN}}}}}}
\newcommand{\vzero}   {\ensuremath{\textrm{V}^{0}}}
\newcommand{\lam}     {\ensuremath{\Lambda}}
\newcommand{\lab}     {\ensuremath{\bar{\Lambda}}}  
\newcommand{\xim}     {\ensuremath{\Xi^{-}}}
\newcommand{\xizero}  {\ensuremath{\Xi^{0}}}
\newcommand{\xip}     {\ensuremath{\bar{\Xi}^{+}}}
\newcommand{\xib}     {\ensuremath{\bar{\Xi}^{0}}}
\newcommand{\sig}     {\ensuremath{\Sigma^{0}}}
\newcommand{\sib}     {\ensuremath{\bar{\Sigma}^{0}}}
\newcommand{\pimin}   {\ensuremath{\pi^-}}
\newcommand{\piplus}  {\ensuremath{\pi^+}}

\newcommand{\pbar}    {\ensuremath{\bar{\textrm{p}}}}
\newcommand{\ommin}   {\ensuremath{\Omega^-}}
\newcommand{\omplus}  {\ensuremath{\bar{\Omega}^+}}           
\newcommand{\pt}      {\ensuremath{p_{\rb{t}}}}
\newcommand{\mt}      {\ensuremath{m_{\rb{t}}}}
\newcommand{\mtmzero} {\ensuremath{m_{\rb{t}} - m_{\rb{0}}}}
\newcommand{\mtavg}   {\ensuremath{\langle m_{\rb{t}} \rangle - m_{\rb{0}}}}
\newcommand{\minv}    {\ensuremath{m_{\rb{inv}}}}
\newcommand{\dedx}    {\ensuremath{\textrm{d}E/\textrm{d}x}}
\newcommand{\dndy}    {\ensuremath{\textrm{d}N/\textrm{d}y}}

\newcommand{\nwound}  {\ensuremath{\langle N_{\rb{w}} \rangle}}
\newcommand{\navg}    {\ensuremath{\langle N \rangle}}
\newcommand{\tf}      {\ensuremath{T_{\rb{f}}}}
\newcommand{\tch}     {\ensuremath{T_{\rb{ch}}}}
\newcommand{\mub}     {\ensuremath{\mu_{\rbt{B}}}}
\newcommand{\gams}    {\ensuremath{\gamma_{\rb{s}}}}
\newcommand{\betat}   {\ensuremath{\beta_{\rb{t}}}}
\newcommand{\btavg}   {\ensuremath{\langle \beta_{\rb{t}} \rangle}}
\newcommand{\betas}   {\ensuremath{\beta_{\rb{s}}}}
\newcommand{\der}     {\ensuremath{\textrm{d}}}
\newcommand{\lamavg}  {\ensuremath{\langle \Lambda \rangle}}
\newcommand{\labavg}  {\ensuremath{\langle \bar{\Lambda} \rangle}}
\newcommand{\ximavg}  {\ensuremath{\langle \Xi^{-} \rangle}}
\newcommand{\xipavg}  {\ensuremath{\langle \bar{\Xi}^{+} \rangle}}

\newcommand{\piavg}   {\ensuremath{\langle \pi \rangle}}
\newcommand{\pipavg}  {\ensuremath{\langle \pi^{+} \rangle}}
\newcommand{\kpavg}   {\ensuremath{\langle \textrm{K}^+ \rangle}}

\newcommand{\ebeam}   {\ensuremath{E_{\rb{beam}}}}
\newcommand{\yproj}   {\ensuremath{y_{\rb{proj}}}}

\begin{document}



\title{Energy dependence of $\Lambda$ and $\Xi$ production in central
Pb+Pb collisions at 20$A$, 30$A$, 40$A$, 80$A$, and 158\agev\
measured at the CERN Super Proton Synchrotron}





%
%

\affiliation{NIKHEF, 
             Amsterdam, Netherlands.}
\affiliation{Department of Physics, University of Athens, 
             Athens, Greece.}
\affiliation{Comenius University, 
             Bratislava, Slovakia.}
\affiliation{KFKI Research Institute for Particle and Nuclear Physics,
             Budapest, Hungary.} 
\affiliation{MIT, 
             Cambridge, USA.}
\affiliation{Henryk Niewodniczanski Institute of Nuclear Physics, 
             Polish Academy of Sciences, 
             Cracow, Poland.}
\affiliation{Gesellschaft f\"{u}r Schwerionenforschung (GSI),
             Darmstadt, Germany.} 
\affiliation{Joint Institute for Nuclear Research, 
             Dubna, Russia.}
\affiliation{Fachbereich Physik der Universit\"{a}t, 
             Frankfurt, Germany.}
\affiliation{CERN, 
             Geneva, Switzerland.}
\affiliation{Institute of Physics \'{S}wi\c{e}tokrzyska Academy, 
             Kielce, Poland.}
\affiliation{Fachbereich Physik der Universit\"{a}t, 
             Marburg, Germany.}
\affiliation{Max-Planck-Institut f\"{u}r Physik, 
             Munich, Germany.}
\affiliation{Charles University, Faculty of Mathematics and Physics,
             Institute of Particle and Nuclear Physics, 
             Prague, Czech Republic.} 
\affiliation{Department of Physics, Pusan National University, 
             Pusan, Republic of Korea.} 
\affiliation{Nuclear Physics Laboratory, University of Washington,
             Seattle, WA, USA.} 
\affiliation{Atomic Physics Department, Sofia University St.~Kliment Ohridski, 
             Sofia, Bulgaria.} 
\affiliation{Institute for Nuclear Research and Nuclear Energy, 
             Sofia, Bulgaria.}
\affiliation{Department of Chemistry, Stony Brook Univ. (SUNYSB), 
             Stony Brook, USA.}
\affiliation{Institute for Nuclear Studies, 
             Warsaw, Poland.}
\affiliation{Institute for Experimental Physics, University of Warsaw,
             Warsaw, Poland.} 
\affiliation{Faculty of Physics, Warsaw University of Technology, 
             Warsaw, Poland.}
\affiliation{Rudjer Boskovic Institute, 
             Zagreb, Croatia.}


%
%

\author{C.~Alt}
\affiliation{Fachbereich Physik der Universit\"{a}t, 
             Frankfurt, Germany.}
\author{T.~Anticic} 
\affiliation{Rudjer Boskovic Institute, 
             Zagreb, Croatia.}
\author{B.~Baatar}
\affiliation{Joint Institute for Nuclear Research, 
             Dubna, Russia.}
\author{D.~Barna}
\affiliation{KFKI Research Institute for Particle and Nuclear Physics,
             Budapest, Hungary.} 
\author{J.~Bartke}
\affiliation{Henryk Niewodniczanski Institute of Nuclear Physics, 
             Polish Academy of Sciences, 
             Cracow, Poland.}
\author{L.~Betev}
\affiliation{CERN, 
             Geneva, Switzerland.}
\author{H.~Bia{\l}\-kowska} 
\affiliation{Institute for Nuclear Studies, 
             Warsaw, Poland.}
\author{C.~Blume}
\affiliation{Fachbereich Physik der Universit\"{a}t, 
             Frankfurt, Germany.}
\author{B.~Boimska}
\affiliation{Institute for Nuclear Studies, 
             Warsaw, Poland.}
\author{M.~Botje}
\affiliation{NIKHEF, 
             Amsterdam, Netherlands.}
% ...



\collaboration{The NA49 Collaboration}
\noaffiliation




\begin{abstract}

Results on \lam, \lab, \xim, and \xip\ production in central Pb+Pb reactions
at 20$A$, 30$A$, 40$A$, 80$A$, and 158\agev\ are presented.  The energy
dependence of transverse mass spectra, rapidity spectra, and multiplicities 
is discussed.  Comparisons to string hadronic models (UrQMD and HSD) and 
statistical hadron gas models are shown.  While the latter provide a 
reasonable description of all particle yields, the first class of models 
fails to match the \xim\ and \xip\ multiplicities.

\end{abstract}




\maketitle

\section{Introduction}

Heavy ion reactions at ultra-relativistic energies allow the study of
strongly interacting matter at extreme temperatures and densities.  It
is expected that under such conditions eventually a quark gluon plasma 
(QGP) will be formed.  In this state of matter the normal confinement 
of quarks and gluons in hadrons is removed and the partons can exist as 
quasi-free particles in an extended region of space-time.  One of the 
first signatures proposed for the formation of a QGP state was an 
enhancement of strange particle production in A+A with respect to p+p 
collisions \cite{RAFELSKI}.  The argumentation relies on the assumption 
that gluon fusion processes, which may be dominant in a QGP, produce 
additional s$\bar{\textrm{s}}$~pairs\cite{KOCH}.  This in turn will cause 
the abundance of strange quarks to reach its chemical equilibrium value 
in much shorter times than would be possible in a pure hadronic scenario.  
In fact, the enhanced production of strange particles has been observed 
already quite early in high energy nucleus--nucleus collisions 
\cite{NA35LAM,NA35STR}.  It has also been demonstrated that the enhancement 
is most pronounced for the multiply strange hyperons $\Xi$ and $\Omega$ 
\cite{WA97HYP,NA57HY158,NA49MICHI}.
 
However, systematic studies of hadron production in nucleus--nucleus
collisions have shown that strangeness enhancement is not only seen
at high energies, such as top SPS and RHIC energies, but it is also 
observed at lower energies ($\sqrts < 5$~GeV) \cite{E802STR} where no 
QGP formation is expected.  Actually, here the production of $\Lambda$ and 
$\Xi$ exhibits an even stronger enhancement than present at top SPS or 
RHIC energies \cite{E895HYP,NA49LAM,NA49OM}.  Generally, it is found that 
the abundances of strange particles in central A+A reactions are similar
to those expected from statistical hadron gas models assuming a grand 
canonical ensemble \cite{BECATTINI1,PBM1}.   While the enhancement at
lower energies can to a certain extent also be explained by transport models,
at higher energies additional mechanisms have to be involved in order
to reach chemical equilibrium values via a dynamical evolution.  It has, 
e.g., been suggested that multi-pion reaction processes can lead to an 
accelerated equilibration of anti-hyperon production in nucleus--nucleus 
collisions \cite{CGREINER}.  Especially at larger densities, as present 
close to the QGP phase boundary, processes like this might drive the 
system quite fast to a chemical equilibrium state \cite{PBM2}.  Still it 
is an open question whether such dynamical explanations are applicable as 
well at lower energies.  On the other hand, it was suggested that particle 
production via strong interaction always follows the maximum entropy 
principle and therefore hadron abundances are naturally close to the outcome
of statistical processes \cite{BECATTINI2,BECATTINI3,HEINZ,STOCK}.  The 
measurement of hyperon production in an energy range below top SPS energy 
($\sqrts < 17.3$~GeV) provides important constraints on both, the statistical 
and transport model approach.  Recent results on kaon production in central
nucleus--nucleus collisions \cite{NA49ONSET} indicate a sharp maximum 
of the ratio $\kpavg / \pipavg$ and a sudden change in the energy dependence 
of the \mtavg\ of pions, kaons, and protons at a beam energy of 30\agev, where
$\mt = \sqrt{\pt^{2} + m_{0}^{2}}$ is the transverse mass, $m_{0}$ the rest
mass and \pt\ the transverse momentum.  These anomalies can be interpreted 
as a signal for the onset of deconfinement \cite{SMES,STEP} and might also 
be visible in the energy dependence of hyperon yields.

%
\begin{table}[tbh]
\caption{\label{tab:datasets}
Summary of the analyzed datasets.  The centrality fraction corresponds 
to the most central part of the total inelastic cross section.  
The Glauber model was used to determine the averaged number of wounded 
nucleons per event \nwound. }
\begin{ruledtabular}
\begin{tabular}{cclccll}
  \multicolumn{1}{l}{\ebeam}  &
  \multicolumn{1}{l}{\sqrts}  & 
  $y_{\rb{cm}}$               & 
  \multicolumn{1}{l}{Cent.}   &
  \nwound                     & 
  Year                        & 
  Statistics                  \\ 
  \multicolumn{1}{l}{(\agev)} &
  \multicolumn{1}{l}{(GeV)}   &
                              &
  \multicolumn{1}{c}{(\%)}    &
                              &
                              &
                              \\
                              \hline
20  &  6.3 & 1.88 &  7 & 349 & 2002 & 350k \\
30  &  7.6 & 2.08 &  7 & 349 & 2002 & 420k \\
40  &  8.7 & 2.22 &  7 & 349 & 1999 & 380k($\Lambda$)/580k($\Xi$) \\
80  & 12.3 & 2.56 &  7 & 349 & 2000 & 300k \\
158 & 17.3 & 2.91 & 10 & 335 & 2000 & 1.2M \\
\end{tabular}
\end{ruledtabular}
\end{table}
%

The data discussed here represent an extension of previously published 
results \cite{NA49XI,NA49LAM,NA49OM} in order to provide a complete 
study of the energy dependence of hyperon production at the CERN-SPS.  
Some of the data discussed here have been presented as preliminary 
before~\cite{NA49MICHI,NA49AGNES,NA49CHRIS}.  However, the data shown in
this publication are the result of a completely new and independent 
analysis, which treats all datasets in a consistent manner.  In particular, 
the results for \lam\ and \lab\ include a correction for the feed-down 
from weak decays, which was not applied in the previous 
publication~\cite{NA49LAM}.


\section{Data analysis}

\subsection{\label{sec:exp} Experimental setup and data sets}

The data were taken with the NA49 large acceptance hadron spectrometer
at the CERN SPS.  A detailed description of the apparatus can be found 
in~\cite{NA49NIM}.  With this detector, tracking is performed by four
large-volume Time Projection Chambers (TPCs) in a wide range of phase
space.  Two of these are positioned inside two superconducting dipole
magnets.  In order to assure a similar detector acceptance for all datasets, 
the magnetic field was chosen proportional to the beam energy.  A 
measurement of the specific energy loss \dedx\ in the TPC gas with a 
typical resolution of 4\% provides particle identification at forward 
rapidities.  Time-of-flight detectors improve the particle identification 
at mid-rapidity.  Central Pb+Pb reactions were selected by imposing an 
upper threshold on the energy measured in the projectile fragmentation 
region.  For this measurement the Zero Degree Calorimeter (ZDC) was used 
which is positioned downstream of the TPCs.  A collimator in front of the 
ZDC assures that the acceptance of the calorimeter matches the phase space 
of the projectile fragments and spectator nucleons. 

%
\begin{figure}[t]
\includegraphics[width=1.0\linewidth]{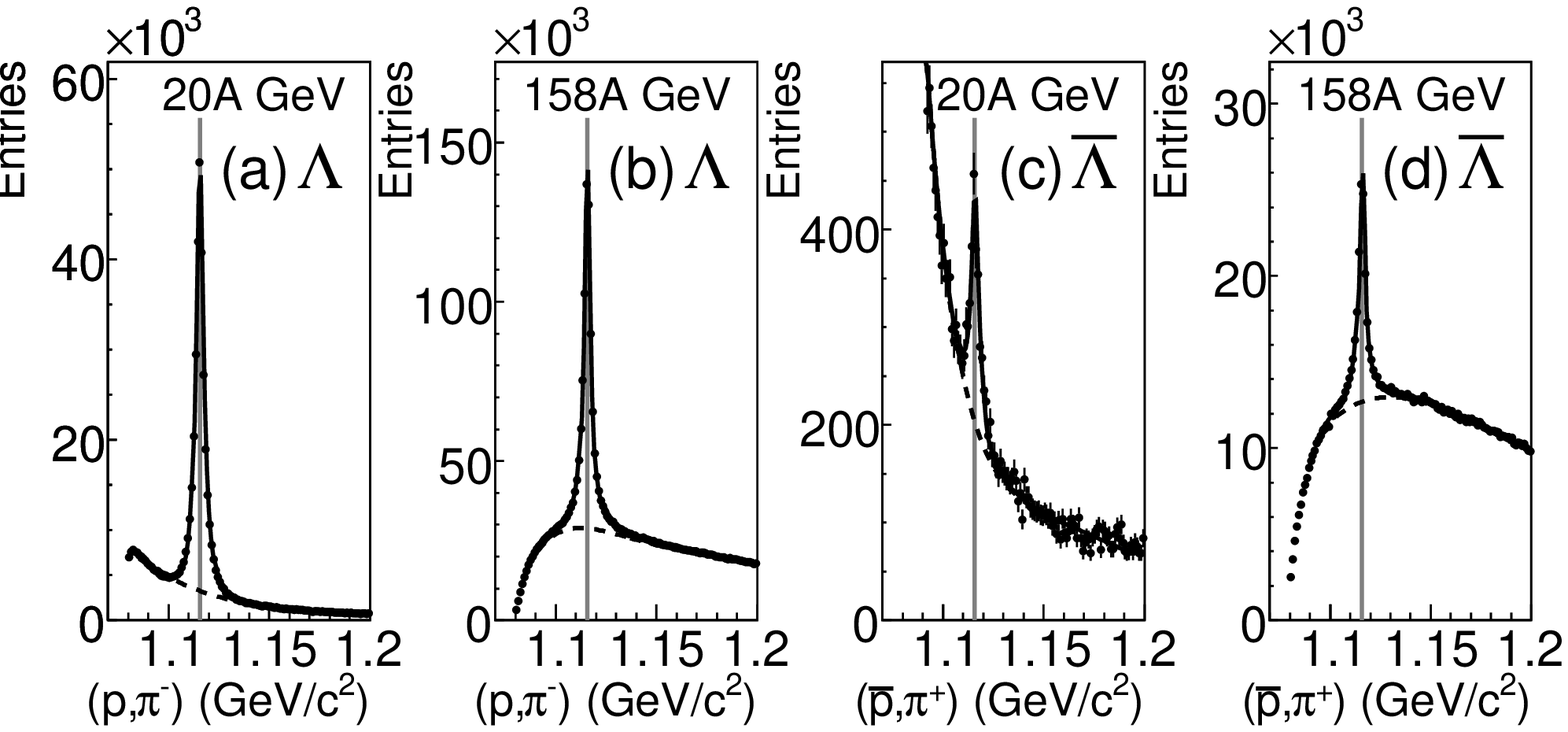}
\includegraphics[width=1.0\linewidth]{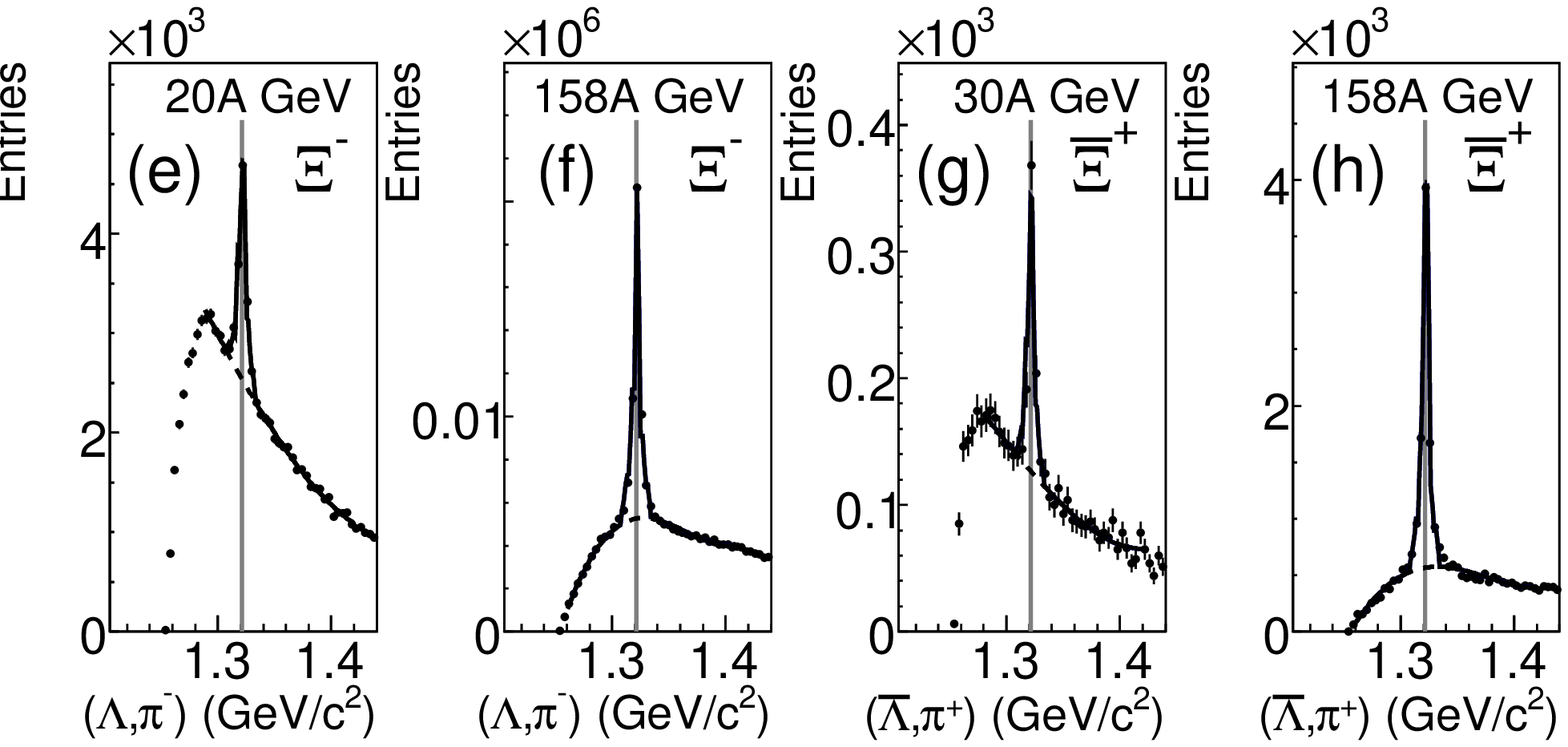}
\caption{The invariant mass distributions of all \lam, \lab\ (upper row), 
\xim, and \xip\ (lower row) candidates in central Pb+Pb collisions at
the lowest and highest analyzed beam energies.  The full curves 
represent a fit to signal and background as described in the text, 
while the dashed curves show the background contribution only.  The
gray vertical lines denote the PDG masses \cite{PDG}.
}
\label{fig:minv_all} 
\end{figure} 
%

%
\begin{figure}[t]
\begin{center}
\begin{minipage}[b]{0.49\linewidth}
\begin{center}
\includegraphics[width=\linewidth]{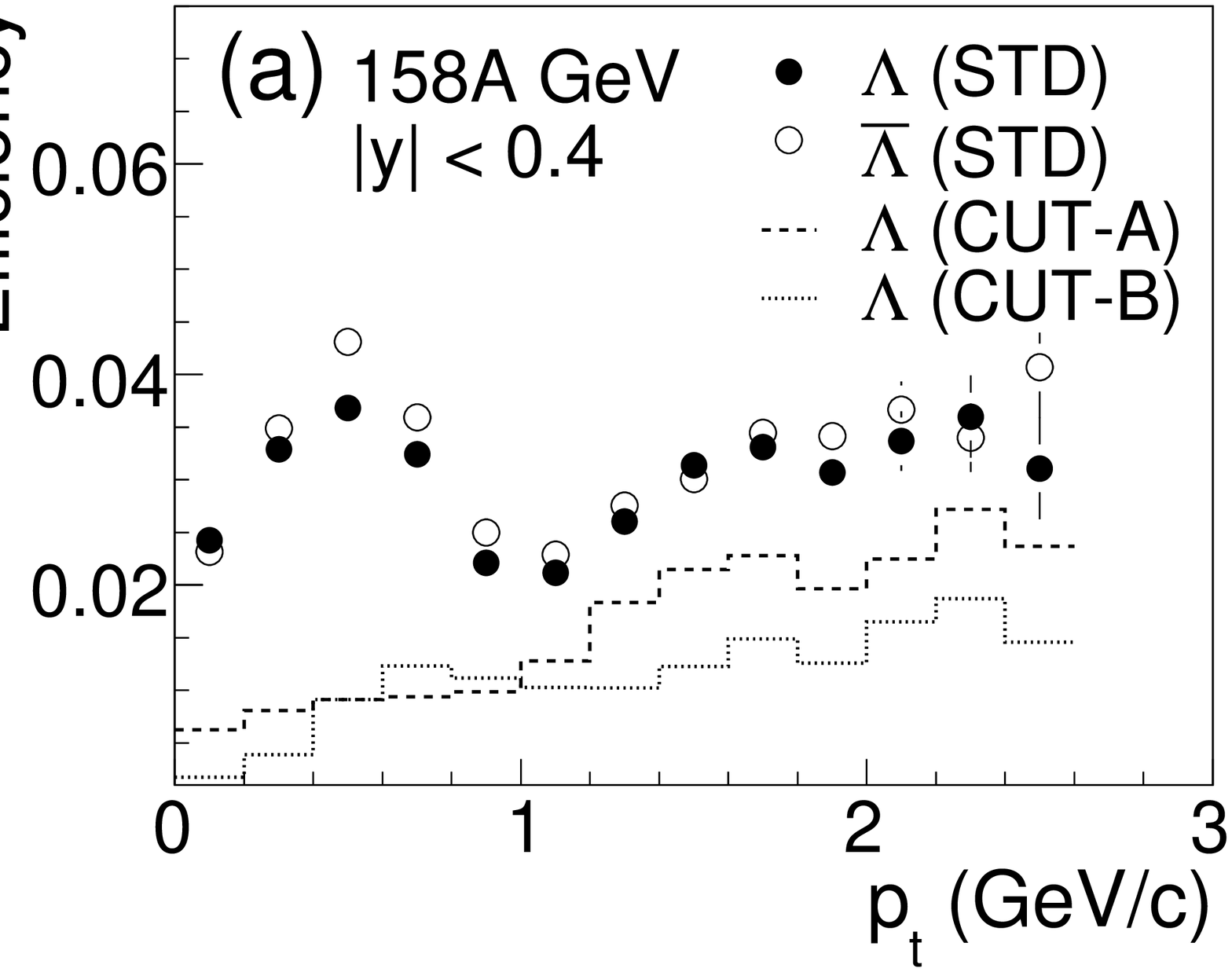}
\end{center}
\end{minipage}
\begin{minipage}[b]{0.49\linewidth}
\begin{center}
\includegraphics[width=\linewidth]{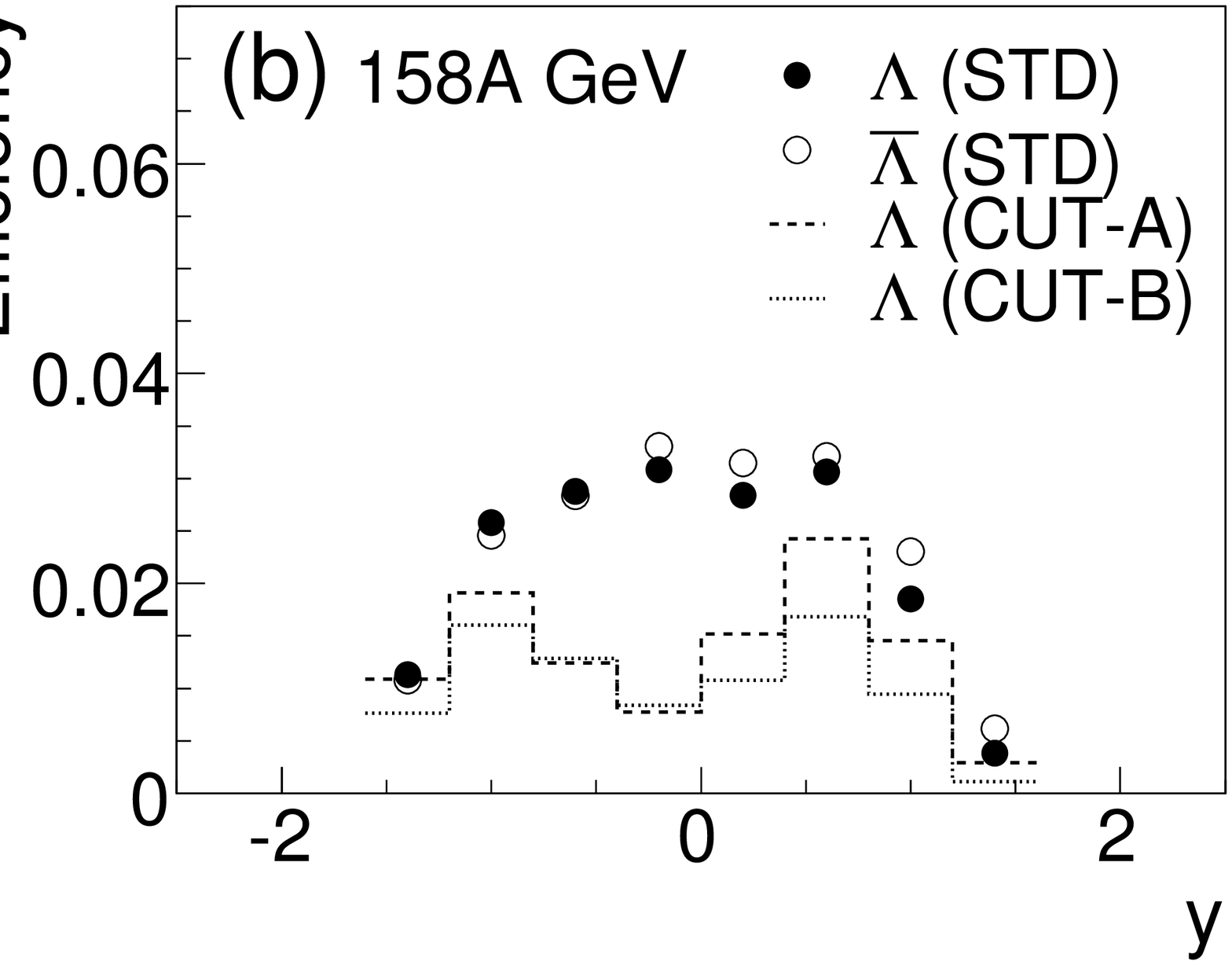}
\end{center}
\end{minipage}
\end{center}
\begin{center}
\begin{minipage}[b]{0.49\linewidth}
\begin{center}
\includegraphics[width=\linewidth]{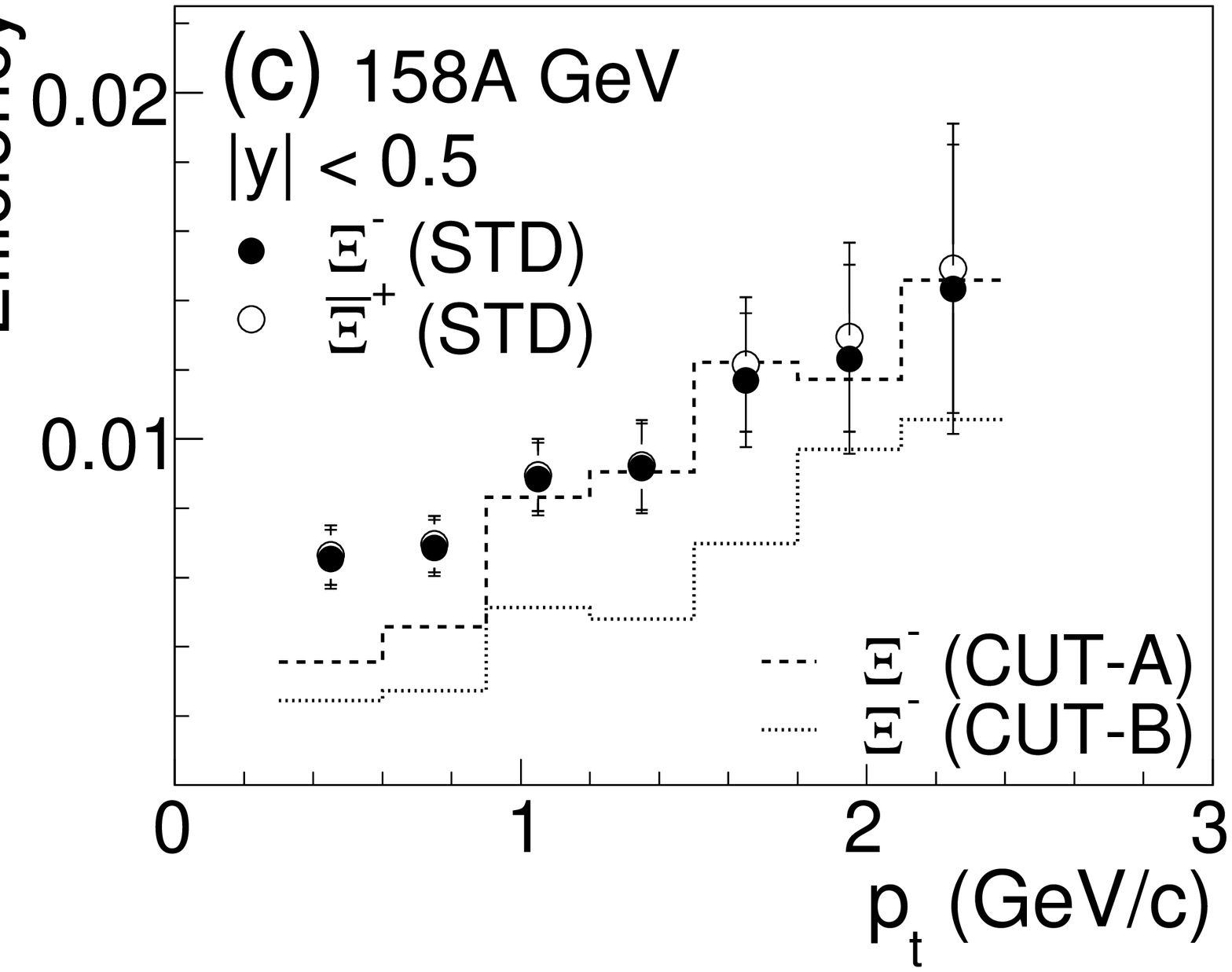}
\end{center}
\end{minipage}
\begin{minipage}[b]{0.49\linewidth}
\begin{center}
\includegraphics[width=\linewidth]{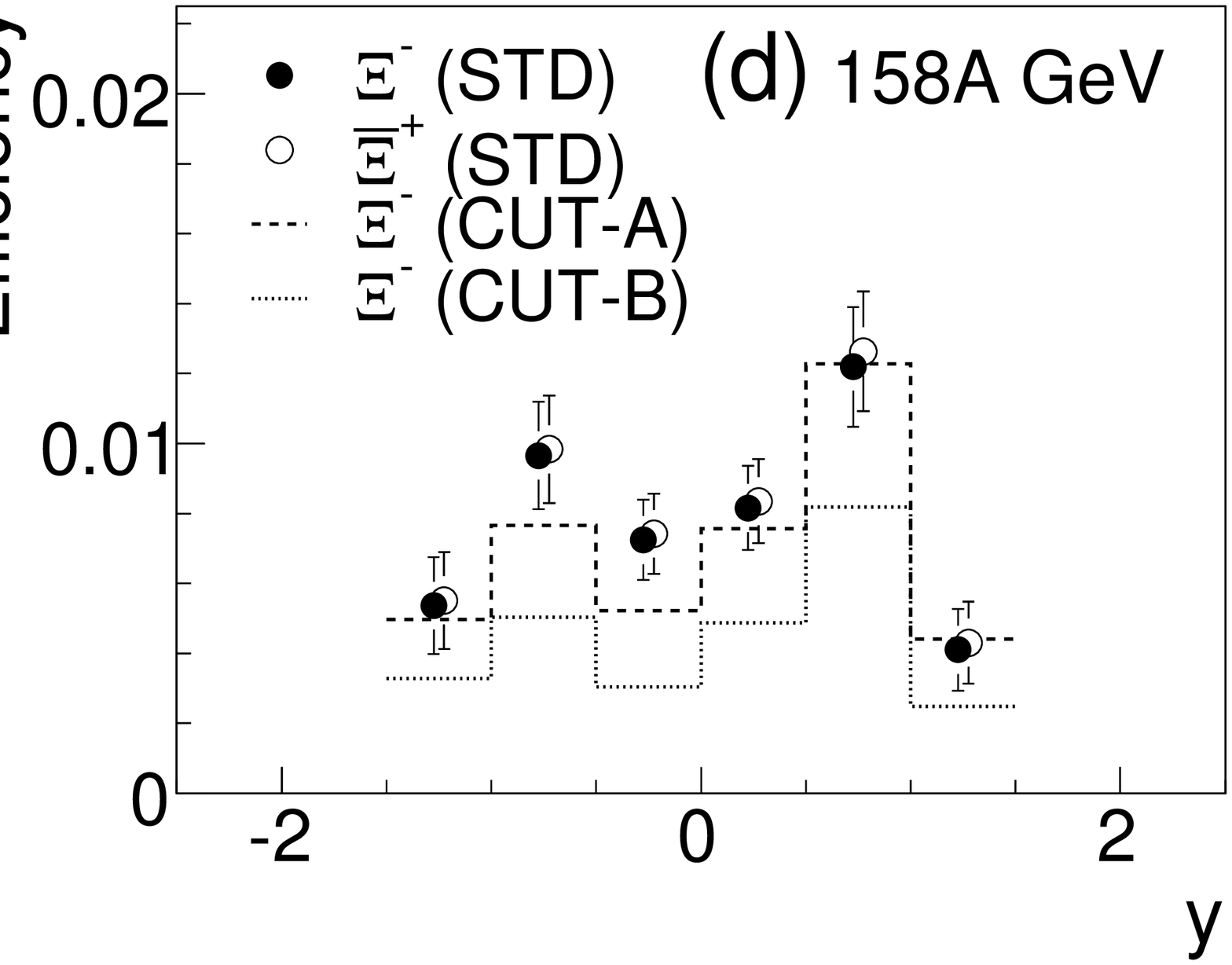}
\end{center}
\end{minipage}
\end{center}
\caption{The total reconstruction efficiency of \lam\ (\lab) (upper
row) and \xim\ (\xip) (lower row) as a function of \pt\ (left panels), 
and as a function of rapidity (right panels) for central Pb+Pb at 
158\agev.  The symbols denote the efficiency for the standard analysis 
procedure (STD).  In addition, the \lam\ (\xim) efficiencies for two 
other selection criteria are shown (dashed: CUT-A, dotted: CUT-B, see 
text).
}
\label{fig:lm_xi_eff} 
\end{figure} 
%

We present in this paper an analysis of central Pb+Pb events taken at 
beam energies of 20$A$, 30$A$, 40$A$, 80$A$, and 158\agev\ in the years 
between 1999~--~2002.  The properties of the different datasets are 
summarized in Table~\ref{tab:datasets}.  The 158\agev\ dataset has an online 
centrality trigger on the 23.5\% most central events, of which the 10\% 
most central were selected offline.

\subsection{\label{sec:rec} \lam\ (\lab) and \xim\ (\xip) reconstruction}

\lam\ and \lab\ hyperons were found by reconstructing their charged decays 
$\lam \rightarrow \pimin + \textrm{p}$ and 
$\lab \rightarrow \piplus + \bar{\textrm{p}}$ (branching ratio 63.9 \% 
\cite{PDG}).  In a first step pairs were formed of all positively charged 
particles with all negatively charged ones.  Their tracks were 
reconstructed by a global tracking algorithm that connects the track parts 
registered in the different TPCs.  Only tracks with more than 10 
reconstructed points were accepted.  By requiring a distance of closest 
approach (DCA) between their trajectories of less than 0.5~cm anywhere 
between the position of the first measured points on the tracks and the 
target plane, \vzero\ candidates were identified.  A set of additional cuts 
was imposed in order to reduce the combinatorial background due to uncorrelated 
pairs.  Identification of (anti-)protons via their specific energy loss (\dedx) 
in the TPCs reduces the contribution of pairs with a wrong mass assignment.  
The measured \dedx\ was required to be within 3.5 standard deviations from 
the predicted Bethe-Bloch value.  A \lam\ (\lab) candidate was accepted if 
the reconstructed position of its decay vertex is at least 30~cm downstream 
from the target and outside the sensitive volume of the TPCs, to avoid 
inefficiencies resulting from an insufficient separation of the clusters of 
the two tracks.  The trajectories of the \lam\ (\lab) candidates were 
extrapolated back to the target plane to determine their impact parameters 
$b_{\rb{x}}$ (in the magnetic bending plane) and $b_{\rb{y}}$ relative to 
the interaction point.  Non-vertex candidates were rejected by requiring 
$|b_{\rb{x}}| < 0.5$~cm and $|b_{\rb{y}}| < 0.25$~cm.  A further reduction 
of the combinatorial background was achieved by placing a requirement on 
the angle $\theta^{*}$ between the flight direction of the mother particle 
and of its positive daughter, determined in the center-of-mass system of the 
\lam\ (\lab) candidate: $-0.95 < \cos \theta^{*} < 0.75$.
Figure~\ref{fig:minv_all}, upper row, shows as examples the resulting 
invariant mass spectra at 20$A$ and 158\agev\ for \lam\ and \lab.  The 
position of the peaks in the \minv~distribution agrees with the nominal 
\lam\ mass determined by the particle data group \cite{PDG}.  From a fit 
with a Gaussian typical mass resolutions of $\sigma_{\rb{m}} \approx 2$~\mevcc\ 
are determined, which depend only slightly on phase space and beam energy.  
Generally, the signal to background ratio ($S/B$) is worse for \lab\ than 
for \lam, due to the lower yield of real \lab.  While $S/B$ decreases with 
energy for \lam, it is constant for \lab.  However, the shape of the 
combinatorial background depends on beam energy in both cases.

\xim\ (\xip) candidates were identified via the decay channel
$\xim \rightarrow \lam + \pimin$ ($\xip \rightarrow \lab + \piplus$)
which has a branching ratio of 99.9 \% \cite{PDG}.  To reconstruct 
the \xim\ (\xip), \lam\ (\lab) candidates were selected in an invariant 
mass window of 1.101~--~1.131~\gevcc\ and combined with all measured 
negatively (positively) charged particles in the event.  The \lam\ (\lab) 
candidates were required to pass the same cuts as described above, with 
the exception of the cuts on $b_{\rb{x/y}}$ and cos~$\theta^{*}$, which were 
not applied here.  The reconstructed \xim\ (\xip) candidates should point 
back to the interaction vertex, while the pions from  \lam\ (\lab) and the 
\xim\ (\xip) decay will on average have a larger impact parameter.
To reject non-vertex candidates, upper limits of $|b_{\rb{x}}| < 0.5$~cm 
and $|b_{\rb{y}}| < 0.25$~cm were therefore imposed on the \xim\ (\xip) 
candidates.  Pions coming from the primary interaction point were 
removed by a cut of $|b_{\rb{y}}| > 1.0$~cm for the negatively 
(positively) charged tracks associated to the \xim\ (\xip) decay 
vertex and $|b_{\rb{y}}| > 0.5$~cm for the negatively (positively) 
charged daughter tracks of the \lam\ (\lab) candidates.  An additional 
improvement of the signal to background ratio was achieved by 
requiring that the $\Lambda$ decay vertex and the pion track were
measured on the same side of the TPCs relative to the beam pipe.  The 
lowest beam energy where a significant \xip~signal could be extracted 
is 30\agev, while \xim\ could be analyzed at all available energies.  
Figure~\ref{fig:minv_all}, lower row, shows the invariant mass 
distributions for \xim\ and \xip\ candidates after all selection 
criteria at the lowest and highest available energies, respectively.  
Similarly as in the case of the \lam\ (\lab) an excellent agreement of 
the peak positions with the PDG masses \cite{PDG} is observed. The 
typical mass resolution, as obtained from a fit with a Gaussian is 
$\sigma_{\rb{m}} \approx 4$~\mevcc.  The dependence of the shape of 
the combinatorial background on the beam energy is less pronounced 
than in the case of \lam\ (\lab).

The invariant mass spectra were fitted to the sum of a polynomial and a 
signal distribution, determined from the simulation procedure described 
below.  The raw yields of \lam, \lab, \xim, and \xip\ are obtained by
subtracting the fitted background and integrating the remaining
signal distributions in a mass window of $\pm 11$~\mevcc\ 
($\pm 10$~\mevcc) around the nominal $\Lambda$ ($\Xi$) mass. 

\subsection{\label{sec:eff} Correction for acceptance and 
reconstruction inefficiency}

Detailed simulations were made to correct the yields for geometrical
acceptance and losses in the reconstruction.  For this purpose, samples 
of $\Lambda$ and $\Xi$ were generated in the full phase space 
accessible to the experiment with \mt~spectra according to:
\begin{equation}
  \label{eq:expo}
  \!\! 
  \frac{1}{\mt} \frac{\der N}{\der \mt \der y} 
  \propto
  \exp \left( - \frac{\mt}{T} \right) .
\end{equation}
and Gaussian shaped distributions in rapidity $y$.  The Geant~3.21 package 
\cite{GEANT3} was used to track the generated particles and their decay
products through the NA49 detector.  Dedicated NA49 software was used 
to simulate the TPC response taking into account all known detector 
effects.  The simulated signals were added to those of real events on 
the raw data level and subjected to the same reconstruction procedure 
as the experimental data.  The acceptances and efficiencies were 
calculated in bins of \pt\ (\mtmzero) and $y$ as the fraction of the 
generated $\Lambda$ ($\Xi$) which traverse the detector, survive the 
reconstruction and pass the analysis cuts.  Of all produced hyperons 
$\approx$~50\% (40\%) of the $\Lambda$ ($\Xi$) appear in the acceptance 
of the detector, i.e. all decay particles are seen in the sensitive 
detector volume.  The reconstruction algorithm together with the cuts 
to suppress the combinatorial background reduce this fraction further 
to $\approx$~6\% (4\%) at 158\agev.  In addition, inefficiencies due 
to the high track multiplicity cause a further reduction.  At 158\agev\ 
this effect is most pronounced and reduces the integrated efficiency 
to $\approx$~2\% (1\%) for $\Lambda$ ($\Xi$).  At lower energies the 
influence of the occupancy is weaker and thus the total efficiency 
increases to $\approx$~4\% (2\%) at 20\agev.  The upper row of 
\Fi{fig:lm_xi_eff} shows the total reconstruction efficiency which 
includes acceptance and all reconstruction inefficiencies for \lam\ 
and \lab\ at the highest beam energy, where the effects of the high 
track density are largest.  
Also included in Fig.~\ref{fig:lm_xi_eff} are efficiencies that have 
been calculated for two analysis strategies different from the default
version described above.  The first one (CUT-A, shown as dashed lines) 
employs a set of selection criteria that depend on the sub-detector
in which a \vzero\ was measured and which were optimized for a large
signal-to-background ratio \cite{NA49XI}.  The second strategy (CUT-B, 
shown as dotted lines) uses the same cuts as described above, but in 
addition only accepts tracks which lie outside the high track density 
region (i.e. 4~cm above or below the middle plane of the TPCs).  This 
criterion allows to minimize the losses due to the high occupancy at 
the expense of a much reduced acceptance in particular at low \pt.  It 
was used in a previous analysis of the \lam\ (\lab) at 158\agev\ 
published in \cite{NA49LAM}.  Both approaches drastically reduce the 
number of reconstructed $\Lambda$ ($\Xi$).  Therefore, they were not 
used as the standard procedure in this analysis, but can serve as a 
cross check that helps to estimate systematic errors (see 
section~\ref{sec:sys_err}).

\subsection{Correction of feed-down to \lam\ (\lab)}

%
\begin{figure}[t]
\begin{center}
\begin{minipage}[b]{0.49\linewidth}
\begin{center}
\includegraphics[width=\linewidth]{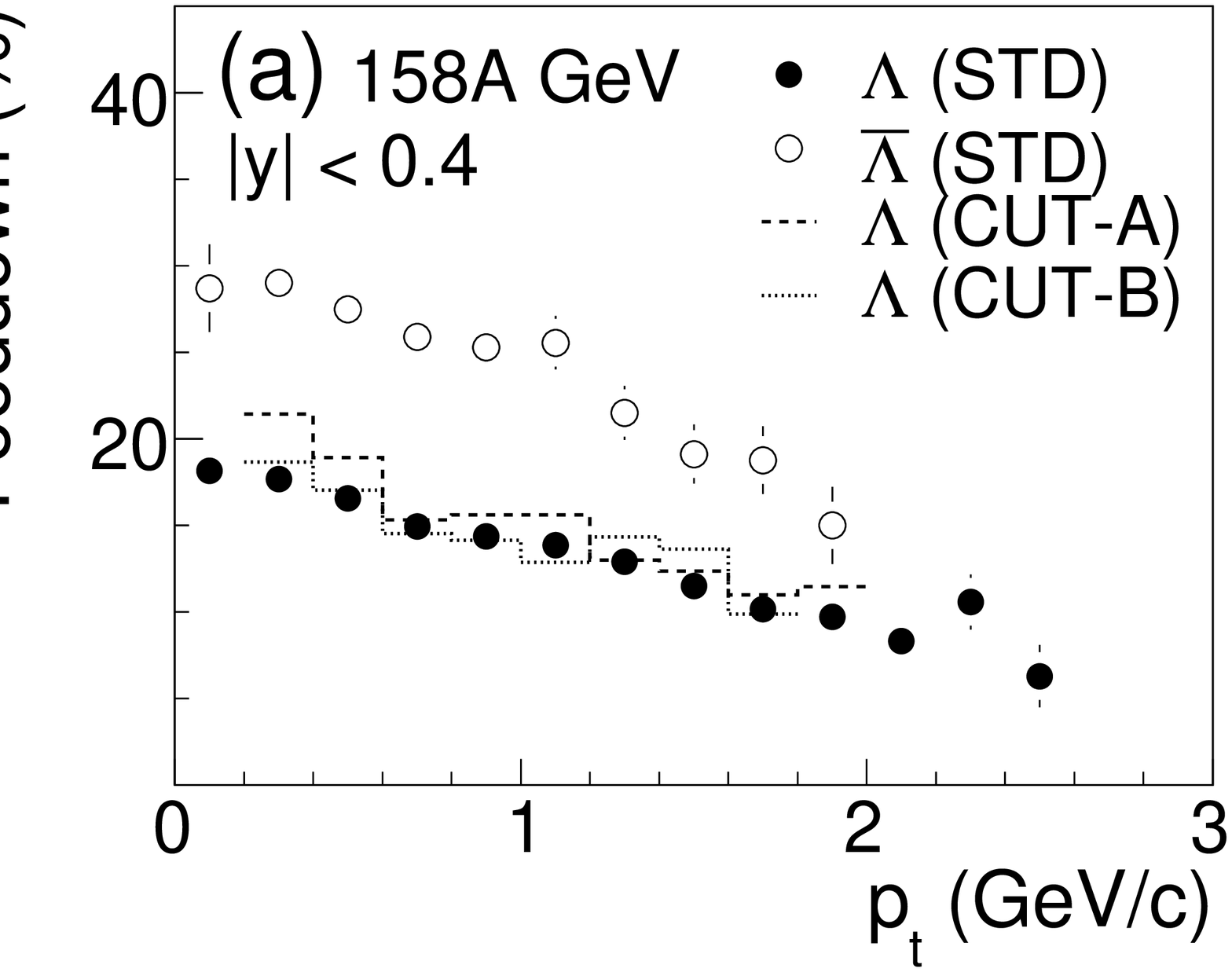}
\end{center}
\end{minipage}
\begin{minipage}[b]{0.49\linewidth}
\begin{center}
\includegraphics[width=\linewidth]{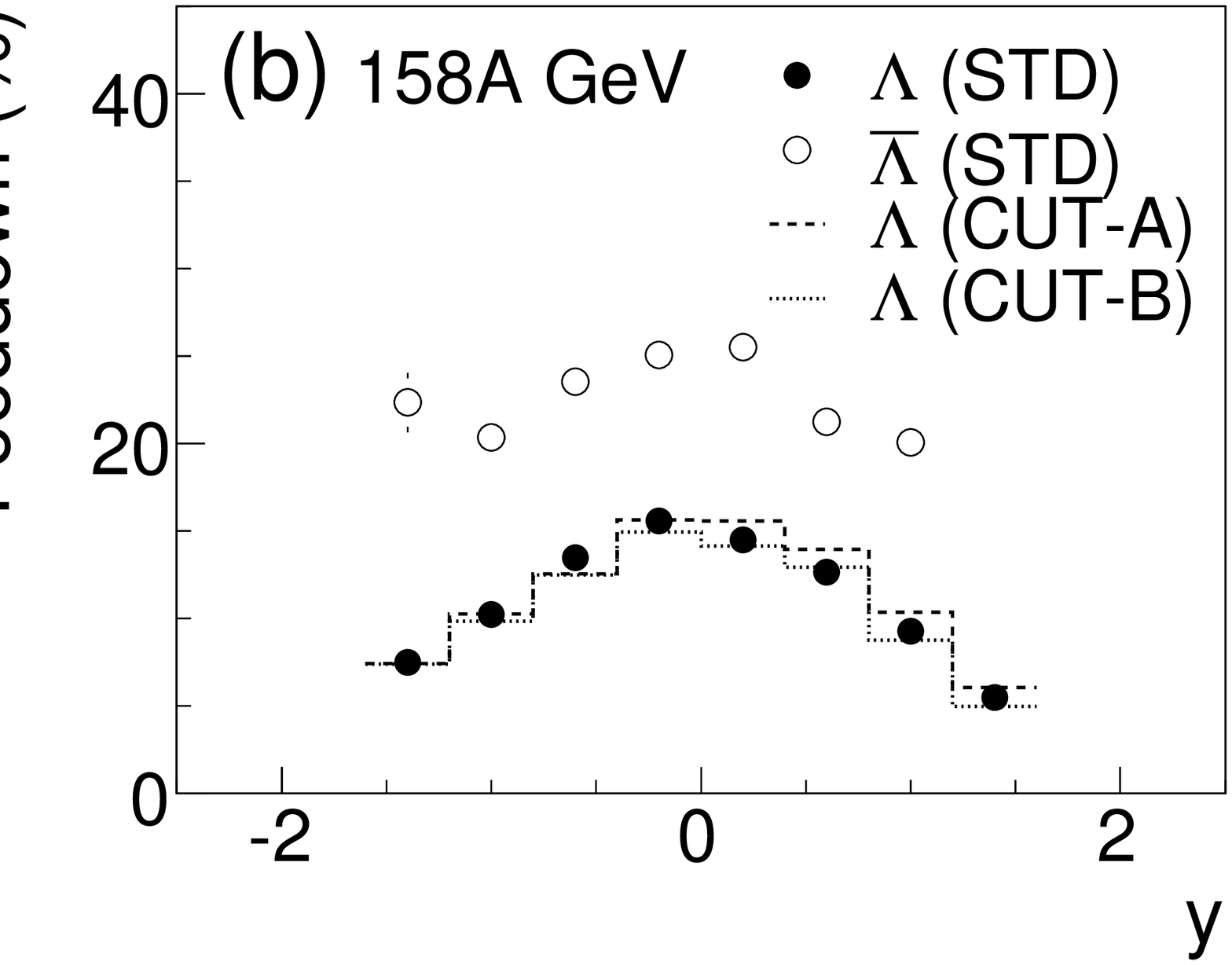}
\end{center}
\end{minipage}
\end{center}
\caption{The feed-down contribution to \lam\ (\lab) around mid-rapidity 
as a function of \pt\ (left panel), and as a function of rapidity 
(right panel) for central Pb+Pb at 158\agev.  The symbols denote the 
feed-down for the standard analysis procedure (STD).  In addition, the 
feed-down to \lam\ for two other selection criteria are shown (dashed: 
CUT-A, dotted: CUT-B, see text).
}
\label{fig:lm_lb_fd} 
\end{figure} 
%

%
\begin{table}[tbh]
\caption{\label{tab:syserr}
Summary of the systematic errors on the \dndy~values.}
\begin{ruledtabular}
\begin{tabular}{lccccc}
     & \multicolumn{1}{l}{Background}
     & \multicolumn{1}{l}{Efficiency}
     & \multicolumn{1}{l}{\pt~Extra-}
     & \multicolumn{1}{l}{Feed-down}
     & \multicolumn{1}{l}{Quadratic}    \\
     & \multicolumn{1}{l}{subtraction}
     & \multicolumn{1}{l}{correction}
     & \multicolumn{1}{l}{polation}
     & \multicolumn{1}{l}{correction}
     & \multicolumn{1}{l}{sum}          \\
                                                  \hline    
\lam &  3\% & 10\% &  --- &  1\% & 10.5\% \\                  
\lab &  3\% & 10\% &  --- &  7\% & 12.5\% \\                  
\xim &  3\% & 10\% &  3\% &  --- & 11\%   \\                  
\xip &  3\% & 10\% &  3\% &  --- & 11\%   \\                  
\end{tabular}
\end{ruledtabular}
\end{table}
%

The measured yield of \lam\ and \lab\ contains, in addition to the 
directly produced particles, contributions from the decay of heavier 
hyperons.  The \lam\ (\lab) resulting from electromagnetic decays of 
\sig\ (\sib) cannot be separated from the directly produced ones.
Thus the here presented yields always represent the sum $\lam + \sig$ 
($\lab + \sib$).  The contribution to \lam\ (\lab) from weak decays, 
however, depends on the chosen analysis cuts, since these decay products 
originate from decay vertices with a sizable distance from the main 
interaction point.  Since the NA49 acceptance for \lam\ (\lab) 
favours those that decay at large distances, the contribution of feed 
down \lam\ (\lab) can be quite substantial.  Therefore, we have 
calculated a correction for the feed-down from $\xim + \xizero$ 
($\xip + \xib$) decays to the measured \lam\ (\lab) sample with the same 
simulation procedure as described above for the efficiency correction.  
In this case a sample of \xim\ and \xizero\ (\xip\ and \xib) was generated 
as input to the NA49 simulation chain.  The feed-down correction was then 
calculated in bins of \pt\ (\mtmzero) and $y$ as the fraction of 
reconstructed \lam\ (\lab) which originate from $\xim + \xizero$ 
($\xip + \xib$) decays and pass the same analysis cuts.  The input 
distributions and yields of the \xim\ (\xip) are the ones measured by NA49 
and presented in this publication.  For the \xizero\ (\xib), which are not 
measured, the same phase space distributions were assumed.  Their yields 
are calculated from the ones of \xim\ (\xip) which are scaled by the 
\xizero/\xim\ (\xib/\xip) ratios taken from a statistical model fit \cite{BECATTINI4}.  
Figure~\ref{fig:lm_lb_fd} shows as an example the calculated feed-down
contribution to \lam\ (\lab) as a function of \pt\ and rapidity.  The 
feed-down is largest at low \pt\ and mid-rapidity and larger for \lab\
(20~--~30~\% at 158\agev) than for \lam\ (5~--~15~\% at 158\agev).  
While for \lab\ no significant dependence of the feed-down on the beam 
energy is observed, the feed-down to \lam\ reduces to 3~--~8~\% at 
20\agev.  Also included in \Fi{fig:lm_lb_fd} are the feed-down 
contributions to \lam\ for the two alternative analysis strategies 
described in the previous section (dashed line: CUT-A, dotted line: 
CUT-B).  Since the fraction of $\Xi$ seen in the reconstructed \lam\ 
sample depends on the selected analysis cuts, the feed-down contribution 
has to be evaluated for each approach separately.

\subsection{\label{sec:sys_err} Systematic errors}

%
\begin{figure}[t]
\begin{center}
\begin{minipage}[b]{0.49\linewidth}
\begin{center}
\includegraphics[width=\linewidth]{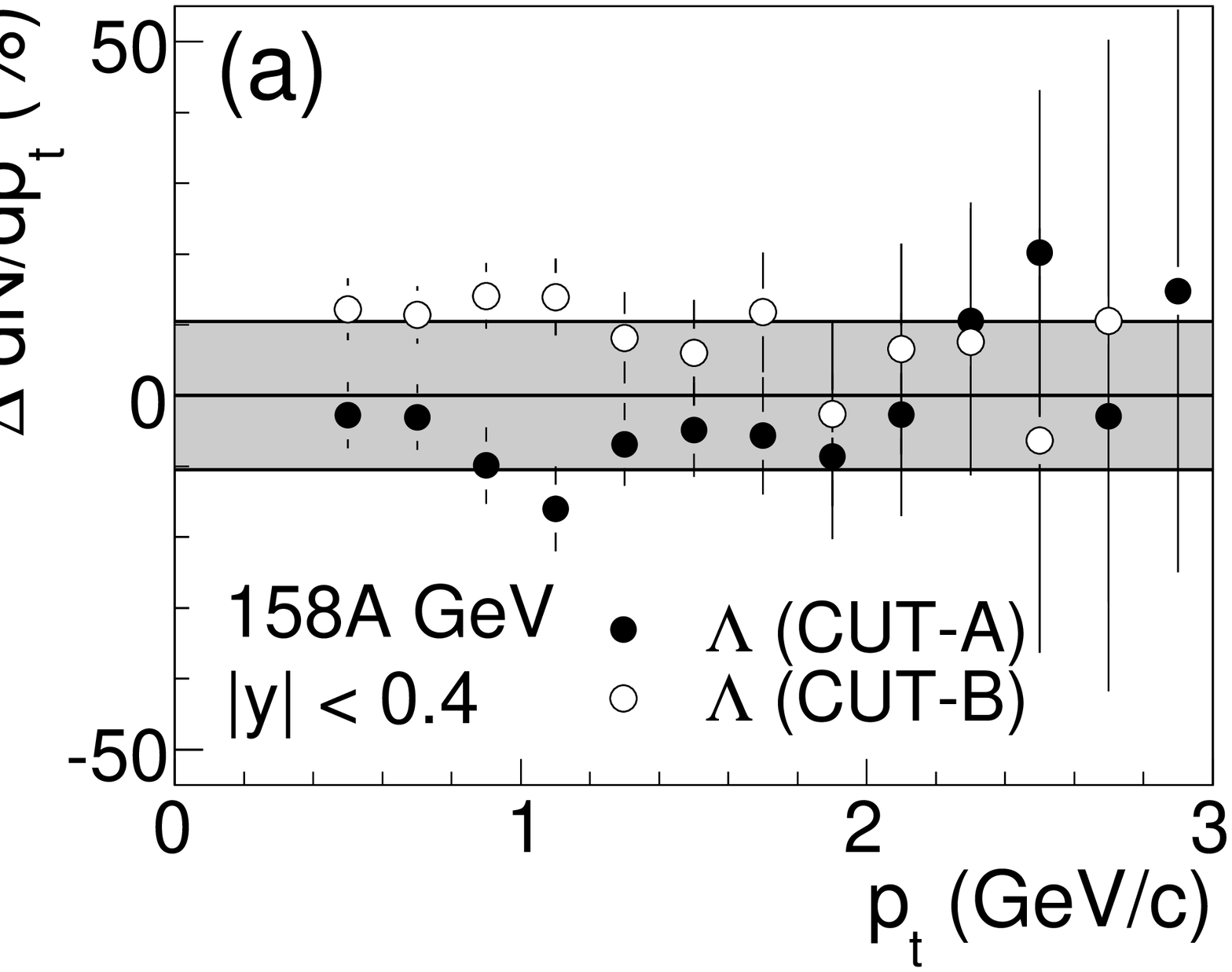}
\end{center}
\end{minipage}
\begin{minipage}[b]{0.49\linewidth}
\begin{center}
\includegraphics[width=\linewidth]{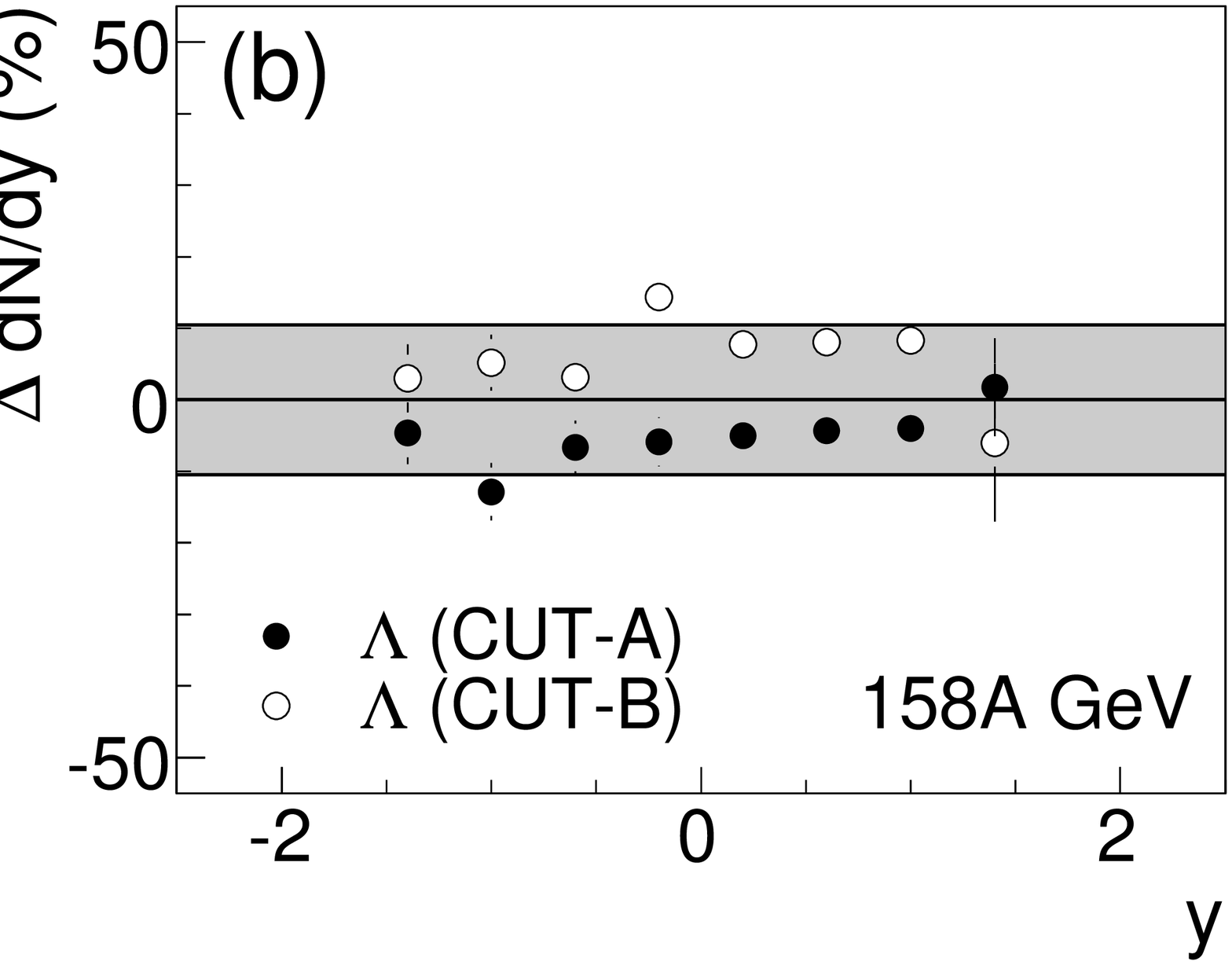}
\end{center}
\end{minipage}
\end{center}
\begin{center}
\begin{minipage}[b]{0.49\linewidth}
\begin{center}
\includegraphics[width=\linewidth]{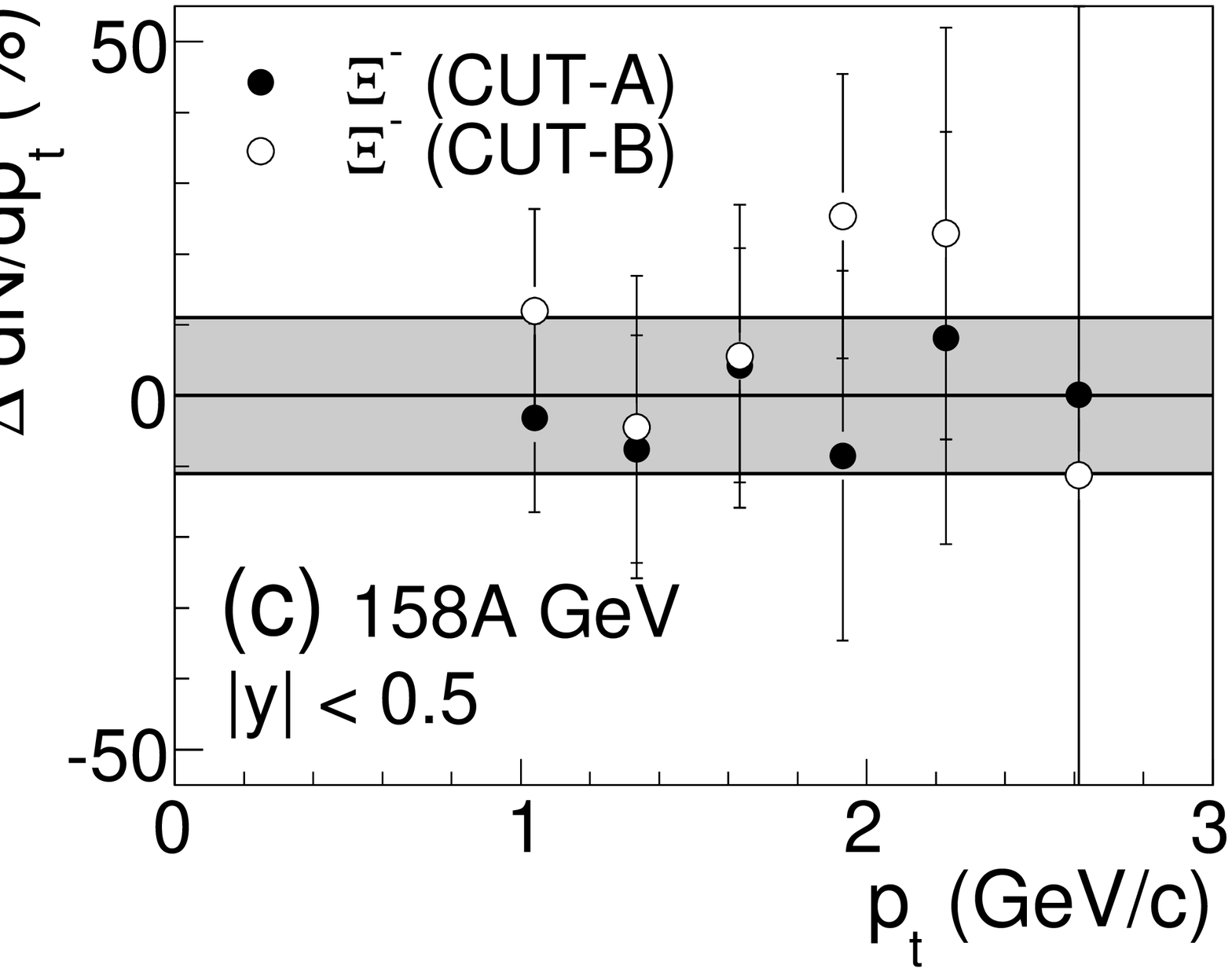}
\end{center}
\end{minipage}
\begin{minipage}[b]{0.49\linewidth}
\begin{center}
\includegraphics[width=\linewidth]{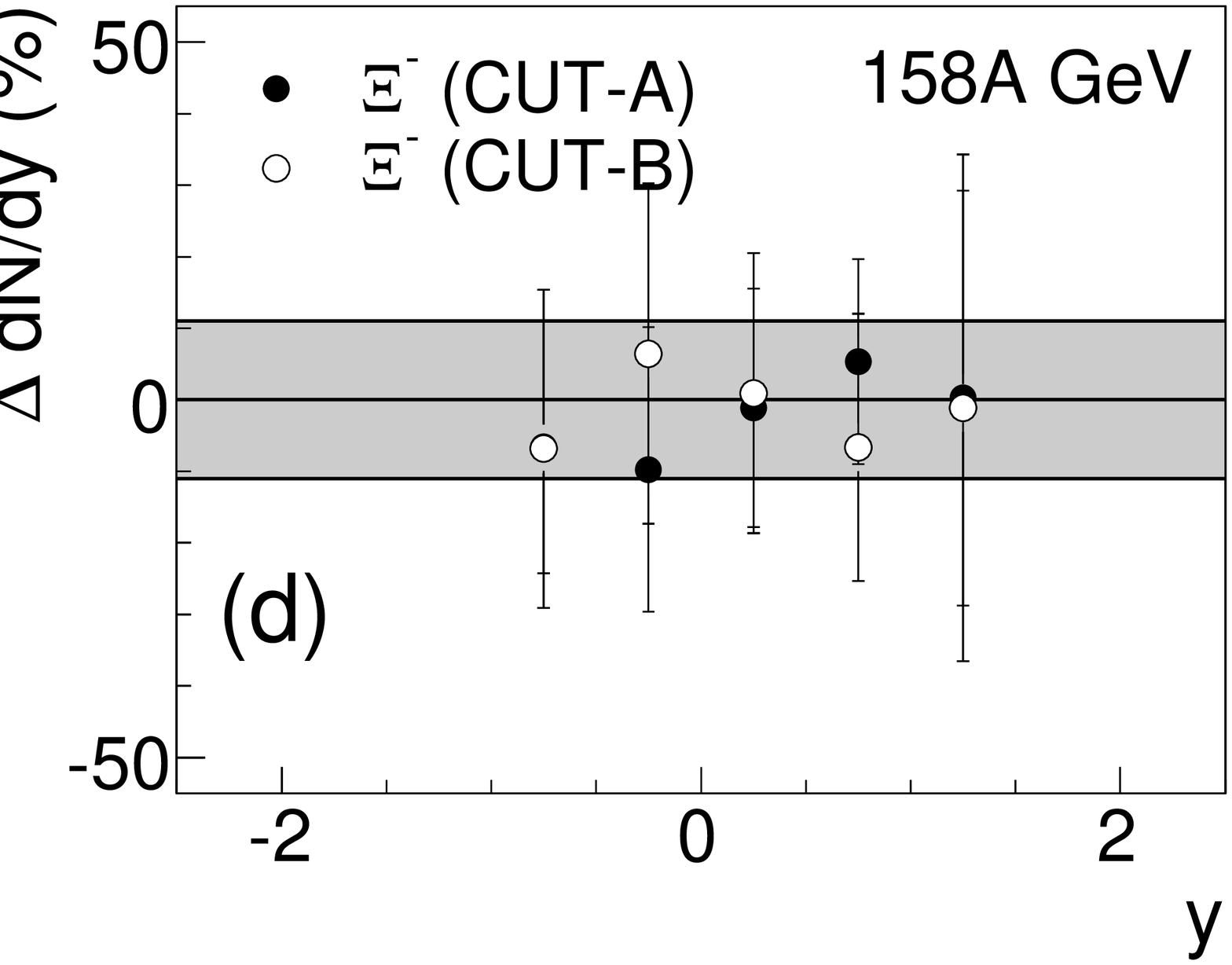}
\end{center}
\end{minipage}
\end{center}
\caption{The differences between the fully corrected results of the 
standard procedure and of the two alternative analysis strategies 
(see section~\ref{sec:sys_err}) for \lam\ (upper row) and \xim\ 
(lower row) in central Pb+Pb at 158\agev.  Shown are the 
\pt~dependence at mid-rapidity (left panels) and the rapidity 
dependence (right panels).  The gray boxes illustrate the systematic 
error estimate.
}
\label{fig:lm_xi_syserr} 
\end{figure} 
%

There are several contributions to the systematic error which are 
summarized in \Ta{tab:syserr}.  One of them results from uncertainties 
in the determination of the combinatorial background.  This uncertainty 
can be estimated by varying the degree of the polynomial used to fit 
the background and the invariant mass range in which the fit is performed.
It is found that this systematic error is 3\% for $\Lambda$ and $\Xi$.  

%
\begin{table*}[tbh]
\caption{\label{tab:summary}
The rapidity densities \dndy\ at mid-rapidity (\lam/\lab: $|y| < 0.4$, 
\xim/\xip: $|y| < 0.5$), the total multiplicities \navg, the RMS-widths 
of the rapidity distributions $RMS_{\rb{y}}$ calculated from the fits
shown in \Fi{fig:all_y}, the averaged transverse masses \mtavg, and the 
inverse slope parameters $T$ at the different beam energies \ebeam.  
The first error is statistical, the second systematic.}
\begin{ruledtabular}
\begin{tabular}{lcclllll}
     & \multicolumn{1}{l}{\ebeam}
     & \multicolumn{1}{l}{Cent.}
     & \dndy
     & \navg
     & $RMS_{\rb{y}}$
     & \mtavg 
     & $T$                                                                                                                         \\
     & \multicolumn{1}{l}{(\agev)}
     & \multicolumn{1}{l}{(\%)}
     & 
     & 
     &
     & (\mevcc)
     & (MeV)                                                                                                                       \\
\hline
\lam & 20  &  7 & 13.4$\pm$0.1$\pm$1.1   & 27.1$\pm$0.2$\pm$2.2   & 0.70$\pm$0.01$\pm$0.06 & 297$\pm$ 4$\pm$24 & 244$\pm$ 3$\pm$12 \\
     & 30  &  7 & 14.7$\pm$0.2$\pm$1.2   & 36.9$\pm$0.3$\pm$3.3   & 0.89$\pm$0.02$\pm$0.08 & 310$\pm$ 5$\pm$25 & 249$\pm$ 2$\pm$13 \\
     & 40  &  7 & 14.6$\pm$0.2$\pm$1.2   & 43.1$\pm$0.4$\pm$4.3   & 1.11$\pm$0.08$\pm$0.11 & 327$\pm$ 5$\pm$27 & 258$\pm$ 3$\pm$13 \\
     & 80  &  7 & 12.9$\pm$0.2$\pm$1.0   & 50.1$\pm$0.6$\pm$5.5   & 1.28$\pm$0.02$\pm$0.14 & 338$\pm$ 7$\pm$27 & 265$\pm$ 4$\pm$13 \\
     & 158 & 10 &  9.5$\pm$0.1$\pm$1.0   & 44.9$\pm$0.6$\pm$8.0   & ---                    & 368$\pm$ 7$\pm$28 & 301$\pm$ 4$\pm$15 \\
\hline
\lab & 20  &  7 & 0.10$\pm$0.02$\pm$0.01 & 0.16$\pm$0.02$\pm$0.03 & 0.62$\pm$0.14$\pm$0.14 & 407$\pm$72$\pm$47 & 339$\pm$56$\pm$31 \\
     & 30  &  7 & 0.21$\pm$0.02$\pm$0.02 & 0.39$\pm$0.02$\pm$0.04 & 0.69$\pm$0.05$\pm$0.08 & 357$\pm$32$\pm$30 & 284$\pm$13$\pm$26 \\
     & 40  &  7 & 0.33$\pm$0.02$\pm$0.03 & 0.68$\pm$0.03$\pm$0.07 & 0.77$\pm$0.05$\pm$0.08 & 371$\pm$22$\pm$31 & 301$\pm$10$\pm$27 \\
     & 80  &  7 & 0.82$\pm$0.03$\pm$0.08 & 1.82$\pm$0.06$\pm$0.19 & 0.83$\pm$0.05$\pm$0.09 & 363$\pm$19$\pm$30 & 292$\pm$10$\pm$27 \\
     & 158 & 10 & 1.24$\pm$0.03$\pm$0.13 & 3.07$\pm$0.06$\pm$0.31 & 1.00$\pm$0.03$\pm$0.09 & 388$\pm$13$\pm$31 & 303$\pm$ 6$\pm$27 \\
\hline
\xim & 20  &  7 & 0.93$\pm$0.13$\pm$0.10 & 1.50$\pm$0.13$\pm$0.17 & 0.64$\pm$0.08$\pm$0.07 & 289$\pm$27$\pm$29 & 221$\pm$14$\pm$13 \\
     & 30  &  7 & 1.17$\pm$0.13$\pm$0.13 & 2.42$\pm$0.19$\pm$0.29 & 0.73$\pm$0.14$\pm$0.09 & 278$\pm$19$\pm$28 & 233$\pm$11$\pm$14 \\
     & 40  &  7 & 1.15$\pm$0.11$\pm$0.13 & 2.96$\pm$0.20$\pm$0.36 & 0.94$\pm$0.13$\pm$0.11 & 285$\pm$17$\pm$29 & 222$\pm$ 9$\pm$13 \\
     & 80  &  7 & 1.22$\pm$0.14$\pm$0.13 & 3.80$\pm$0.26$\pm$0.61 & 0.98$\pm$0.25$\pm$0.16 & 317$\pm$22$\pm$32 & 227$\pm$14$\pm$14 \\
     & 158 & 10 & 1.44$\pm$0.10$\pm$0.15 & 4.04$\pm$0.16$\pm$0.57 & 1.18$\pm$0.18$\pm$0.17 & 327$\pm$13$\pm$33 & 277$\pm$ 9$\pm$17 \\
\hline
\xip & 20  &  7 & ---                    & ---                    & ---                    & ---               & ---               \\
     & 30  &  7 & 0.05$\pm$0.01$\pm$0.01 & 0.12$\pm$0.02$\pm$0.03 & 0.76$\pm$0.35$\pm$0.17 & 326$\pm$60$\pm$33 & 311$\pm$75$\pm$31 \\
     & 40  &  7 & 0.07$\pm$0.01$\pm$0.01 & 0.13$\pm$0.01$\pm$0.02 & 0.65$\pm$0.13$\pm$0.09 & 337$\pm$36$\pm$34 & 277$\pm$32$\pm$28 \\
     & 80  &  7 & 0.21$\pm$0.03$\pm$0.02 & 0.58$\pm$0.06$\pm$0.13 & 0.87$\pm$0.29$\pm$0.20 & 298$\pm$38$\pm$30 & 255$\pm$23$\pm$26 \\
     & 158 & 10 & 0.31$\pm$0.03$\pm$0.03 & 0.66$\pm$0.04$\pm$0.08 & 0.73$\pm$0.08$\pm$0.09 & 384$\pm$26$\pm$38 & 321$\pm$15$\pm$32 \\
\end{tabular}
\end{ruledtabular}
\end{table*}
%

%
\begin{figure*}[t]
\begin{center}
\begin{minipage}[b]{0.48\linewidth}
\begin{center}
\includegraphics[width=60mm]{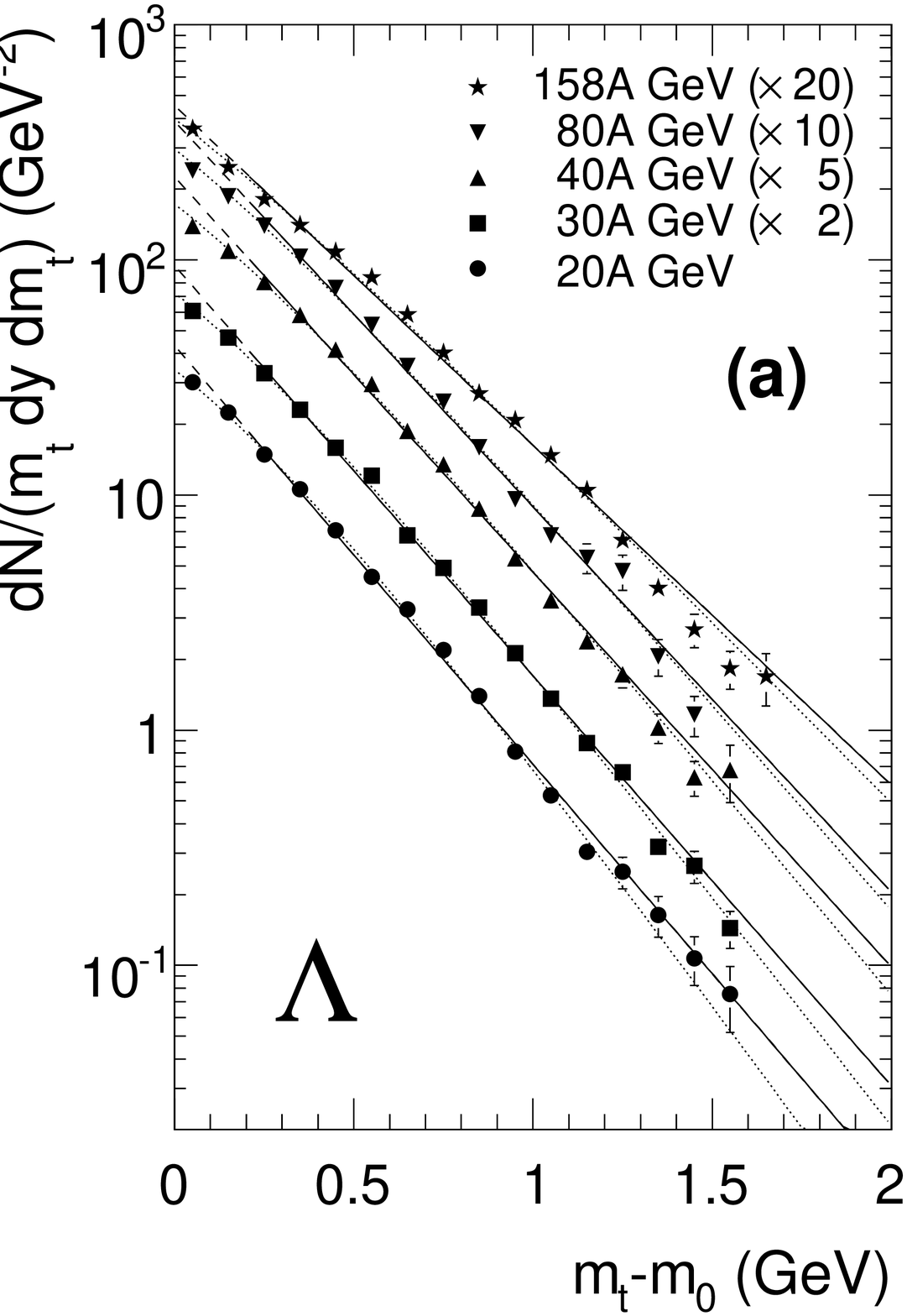}
\end{center}
\end{minipage}
\begin{minipage}[b]{0.48\linewidth}
\begin{center}
\includegraphics[width=60mm]{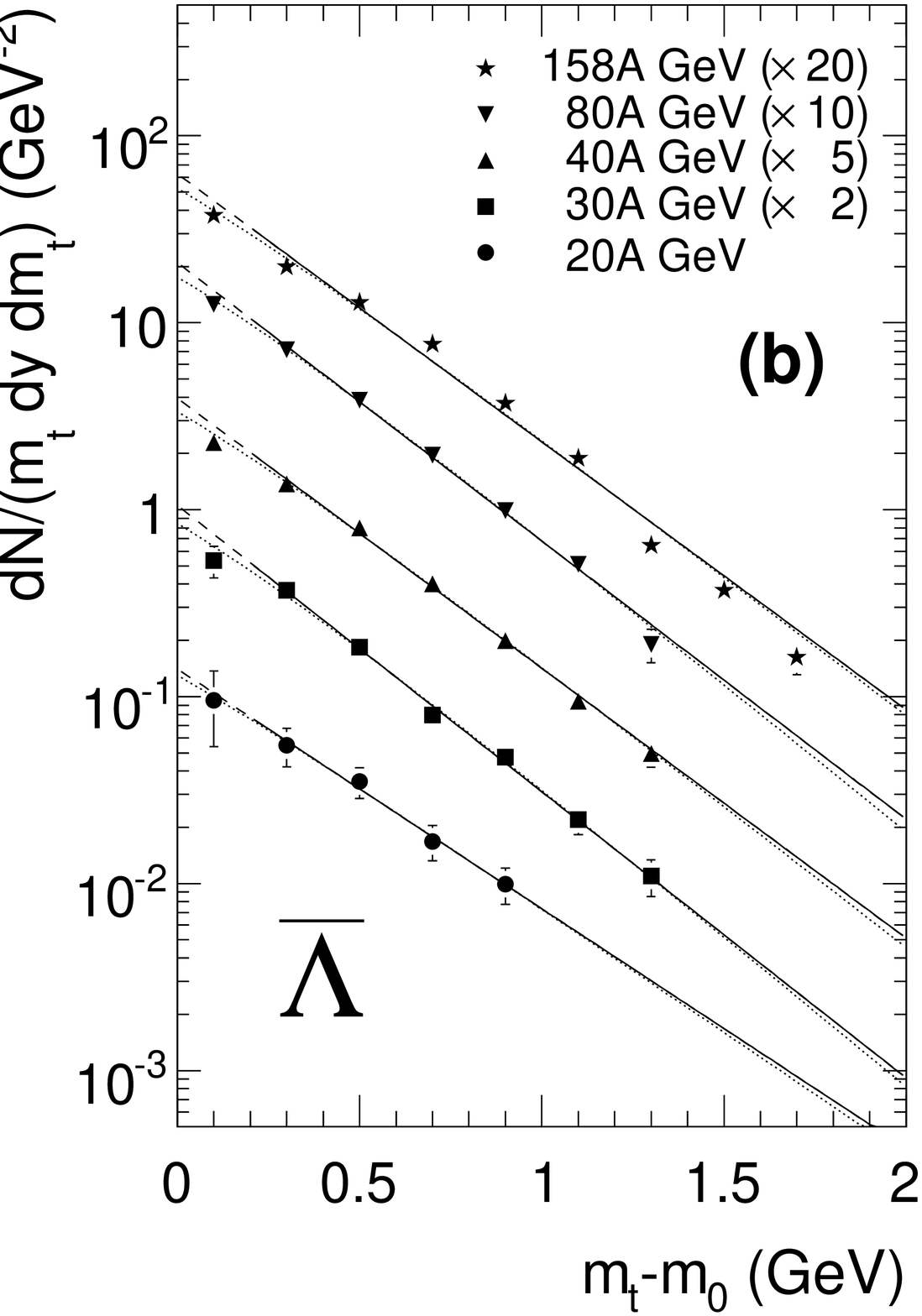}
\end{center}
\end{minipage}
\end{center}
\caption{\label{fig:lm_lb_midrap_mt} 
The transverse mass spectra of \lam\ (left panel) and \lab\ (right panel)
at mid-rapidity ($|y| < 0.4$) for 5 different beam energies. The data
points are scaled for clarity. Only statistical errors are shown.  The 
solid/dashed lines represent a fit with an exponential, where the solid 
part denotes the \mt~range in which the fit was performed.  The dotted 
lines are the results of a fit with a blast wave model \cite{BLASTWAVE}
(see text for details).
}
\end{figure*}
%

Another contribution arises from imperfections in the description of 
the detector response by the simulation procedure which result in 
systematic uncertainties in the efficiency calculation.  It was 
verified that all distributions of geometrical and kinematical 
parameters that are relevant in the reconstruction procedure (see 
section~\ref{sec:rec}) are in agreement between simulated and measured 
data \cite{PHDMICHI,DIPLAGNES,DIPLCHRIS}.  Still there can be remaining 
discrepancies which constitute a source of systematic error.  Its 
magnitude can be estimated by varying the selection criteria in the 
analysis procedure and checking the consistency of the final result. 
This was done, e.g., by comparing data points obtained with the alternative 
analysis strategies described in section~\ref{sec:eff} (CUT-A and CUT-B)
to the results for the standard analysis (see \Fi{fig:lm_xi_syserr}). 
Shown are the differences $\Delta N = N(\rm{STD}) - \it{N}$(CUT-A(B)) 
as a function of \pt\ and rapidity, both for \lam\ and \xim.  Even 
though the efficiencies are lower by almost a factor 2 in some regions 
of phase space (see \Fi{fig:lm_xi_eff}) and are subject to different 
systematic effects (e.g. influence of high track density or background) 
the results are in agreement.  The deviations are consistent with a 
systematic error of $\approx$~10\% for \lam\ and \xim\ at all beam energies 
(see \Fi{fig:lm_xi_syserr}).  Additionally to the studies presented in 
\Fi{fig:lm_xi_syserr}, a further investigation was performed in order to 
test whether the \lam\ reconstruction is sensitive to the cut applied to the 
DCA.  For this purpose the DCA-cut was relaxed to 1.5~cm (default: 0.5~cm) 
and the result of this analysis was compared to the standard procedure.  It
was found that the deviations between the two approaches also agree with 
the systematic error estimate given in \Ta{tab:syserr}.

In case of \lam\ and \lab\ also the uncertainties in the feed-down 
contribution have to be taken into account.  Here, the errors of the 
measurements of spectra and yields of \xim\ and \xip\ translate into 
a systematic error caused by the feed-down correction.  For \lam\ this 
error is small (1\%), since the correction itself is not too substantial 
and the \xim\ measurement is relatively accurate.  In case of \lab, 
however, the larger feed-down contribution and the larger errors of the 
\xip\ data also result in a larger systematic error of 7\%. 

In the range of the errors the data presented here agree well with the 
previously published results where available \cite{NA49LAM,NA49XI}.
The differences compared to the \lam\ and \lab\ yields given in 
\cite{NA49LAM} are due to the feed-down contribution which has not been
subtracted from the old results.

%
\begin{figure*}[t]
\begin{center}
\begin{minipage}[b]{0.48\linewidth}
\begin{center}
\includegraphics[width=60mm]{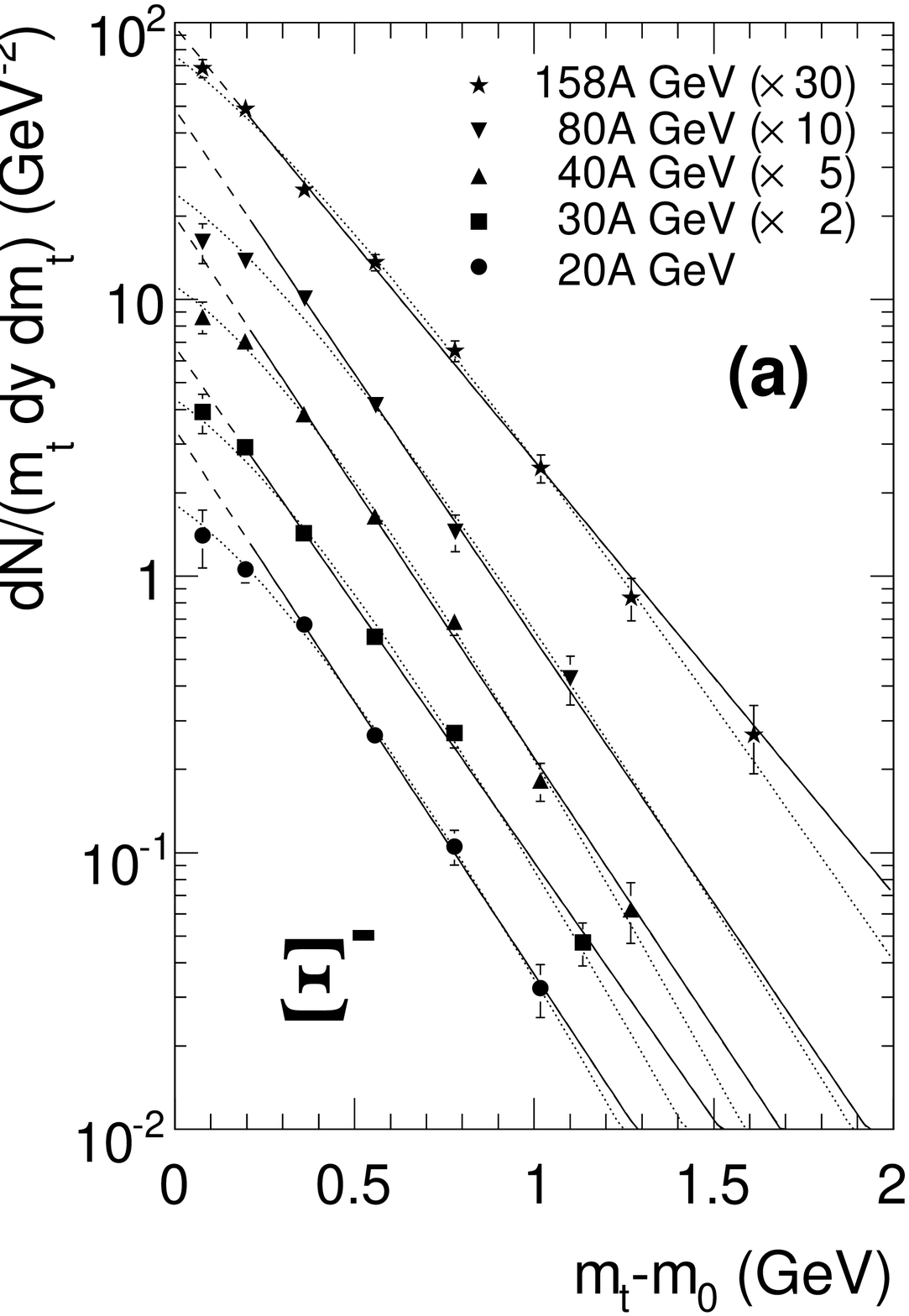}
\end{center}
\end{minipage}
\begin{minipage}[b]{0.48\linewidth}
\begin{center}
\includegraphics[width=60mm]{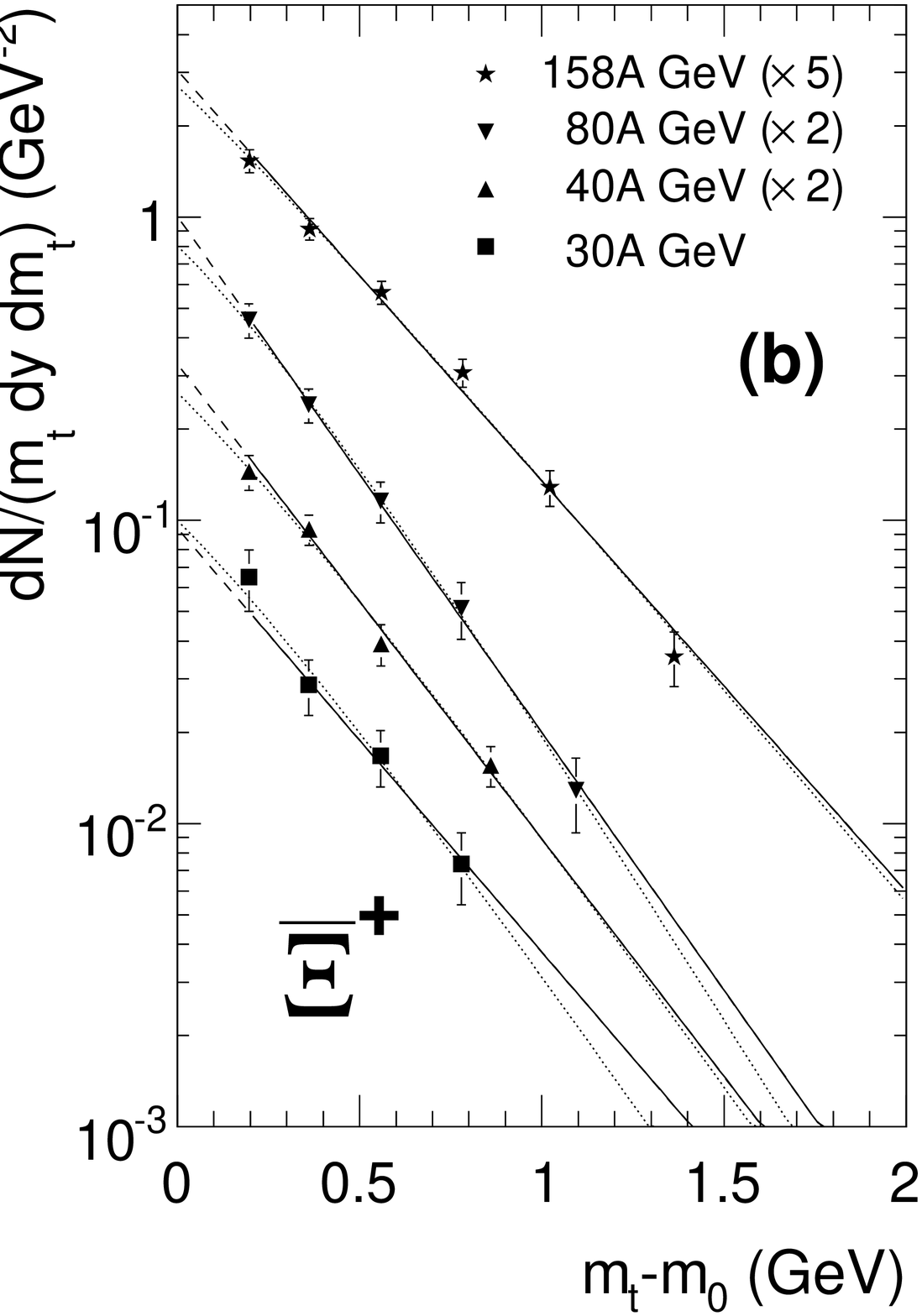}
\end{center}
\end{minipage}
\end{center}
\caption{\label{fig:xi_xb_midrap_mt} 
The transverse mass spectra of \xim\ (left panel) and \xip\ (right panel)
at mid-rapidity ($|y| < 0.5$) for 5 (4) different beam energies. The data
points are scaled for clarity. Only statistical errors are shown.  The 
solid/dashed lines represent a fit with an exponential, where the solid 
part denotes the \mt~range in which the fit was performed.  The dotted 
lines are the results of a fit with a blast wave model \cite{BLASTWAVE}
(see text for details).
}
\end{figure*}
%

Since for the \lam\ (\lab) the range down to $\pt = 0$~\gevc\ is measured
in most of the rapidity bins, the systematic effects due to 
extrapolations into unmeasured \pt~regions is negligible.  Only
in those $y$~bins where extrapolations are necessary an additional
systematic error of 4\% is added in quadrature.  However, for the \xim\ 
(\xip) analysis, this is introducing an additional systematic error in
the full range of the rapidity distributions.  It was estimated by using 
different assumptions for the spectral shape.  The standard approach (fit 
to an exponential, see section~\ref{sec:rap_spectra}) was compared to a fit 
with a hydrodynamically inspired blast wave model~\cite{BLASTWAVE}.  The 
difference on the \dndy\ was found to be 3\%.

\clearpage

The extraction of the total multiplicities requires in addition an 
extrapolation into the unmeasured rapidity regions.  The systematic error 
that is introduced by this extrapolation depends on the beam energy, since 
the fractions of the longitudinal phase space covered by the measurements 
also change with energy.  Also, the shape of the $y$~spectra is not always 
very well determined, especially for \lam\ at 80$A$ and 158\agev.  By using 
different assumptions for the spectral shape in the unmeasured region, as 
defined in \Eq{eq:expo} and \Eq{eq:blast}, the additional systematic error 
on the total multiplicities was estimated.  For \lam\ a variation between 
1\% at 20\agev\ and 14\% at 158\agev\ was obtained and for \lab\ this 
systematic error is largest at the lowest energy (20\%) and decreases to 
2\% at 158\agev.  In case of the \xim\ this contribution ranges between 2\% 
at 20\agev\ and 12\% at 80\agev, while for the \xip\ it is between 5\% 
(158\agev) and 20\% (30$A$ and 80\agev).

\section{Transverse mass spectra}

%
\begin{figure}[ht]
\includegraphics[width=0.8\linewidth]{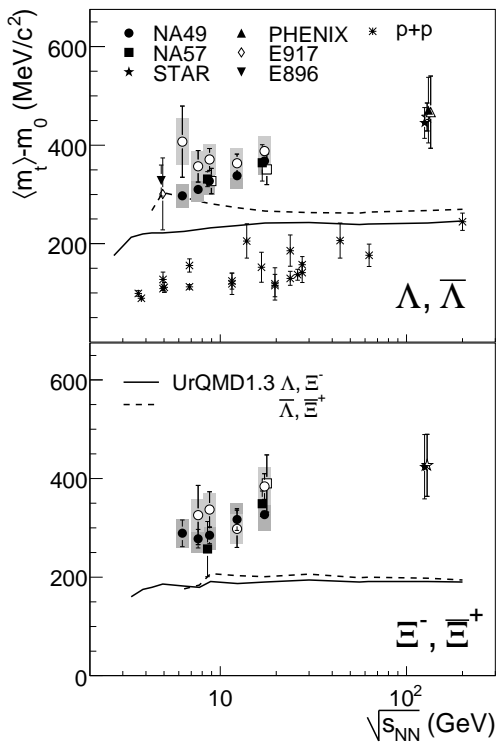}
\caption{The \mtavg\ values for central Pb+Pb and Au+Au reactions as a 
function of \sqrts.  The systematic errors are represented by the 
gray boxes.  Filled symbols correspond to \lam\ and \xim, while open 
symbols denote \lab\ and \xip.  Also shown are data from the NA57 
collaboration \cite{NA57HY40,NA57HY158}, from AGS \cite{E896LAM,E917LAB}
and RHIC experiments \cite{STARLM130,STARXI130,PHNXLM130}, as well 
as p+p data on \lam\ \cite{PPLAM}.  The lines are calculations with 
the UrQMD1.3 model \cite{URQMD,HSDURQMD}.}
\label{fig:meanmt} 
\end{figure} 
%

%
\begin{figure*}[t]
\begin{minipage}[b]{0.95\linewidth}
\includegraphics[width=0.95\linewidth]{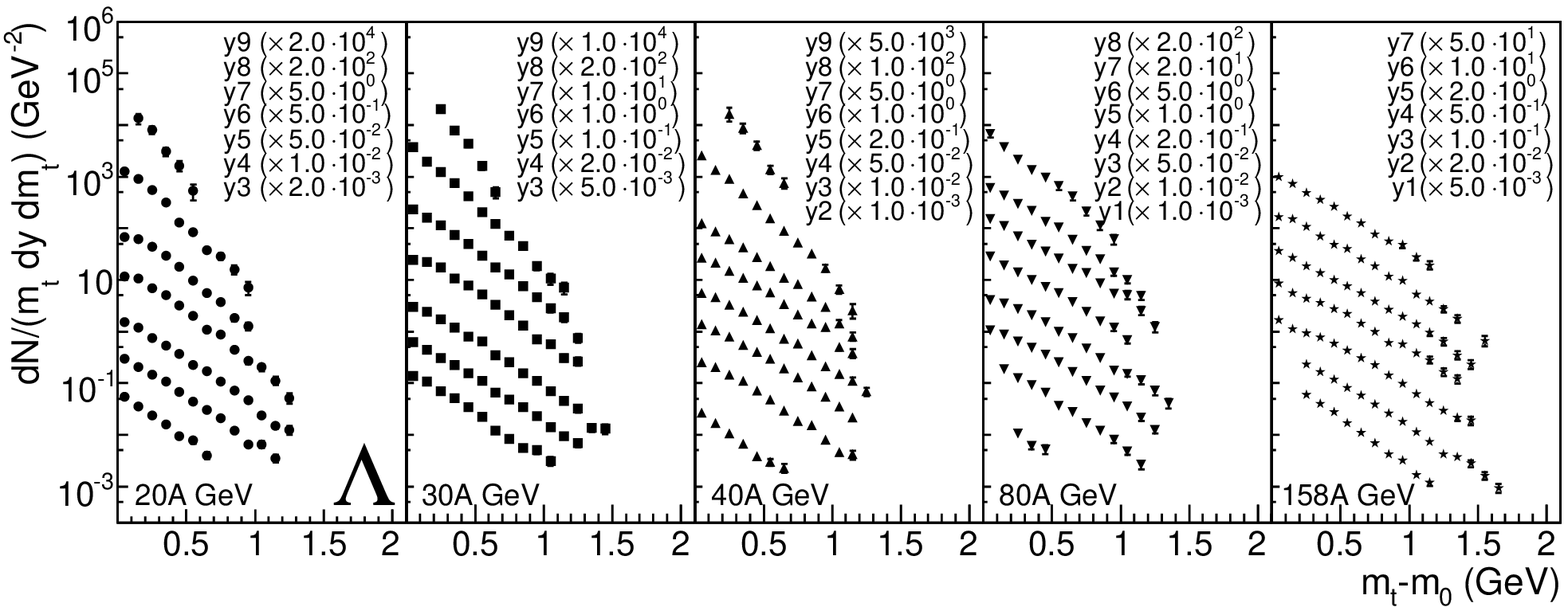}
\end{minipage}
\vspace*{-3.7mm}
\vspace*{-4.8mm}
\begin{minipage}[b]{0.95\linewidth}
\includegraphics[width=0.95\linewidth]{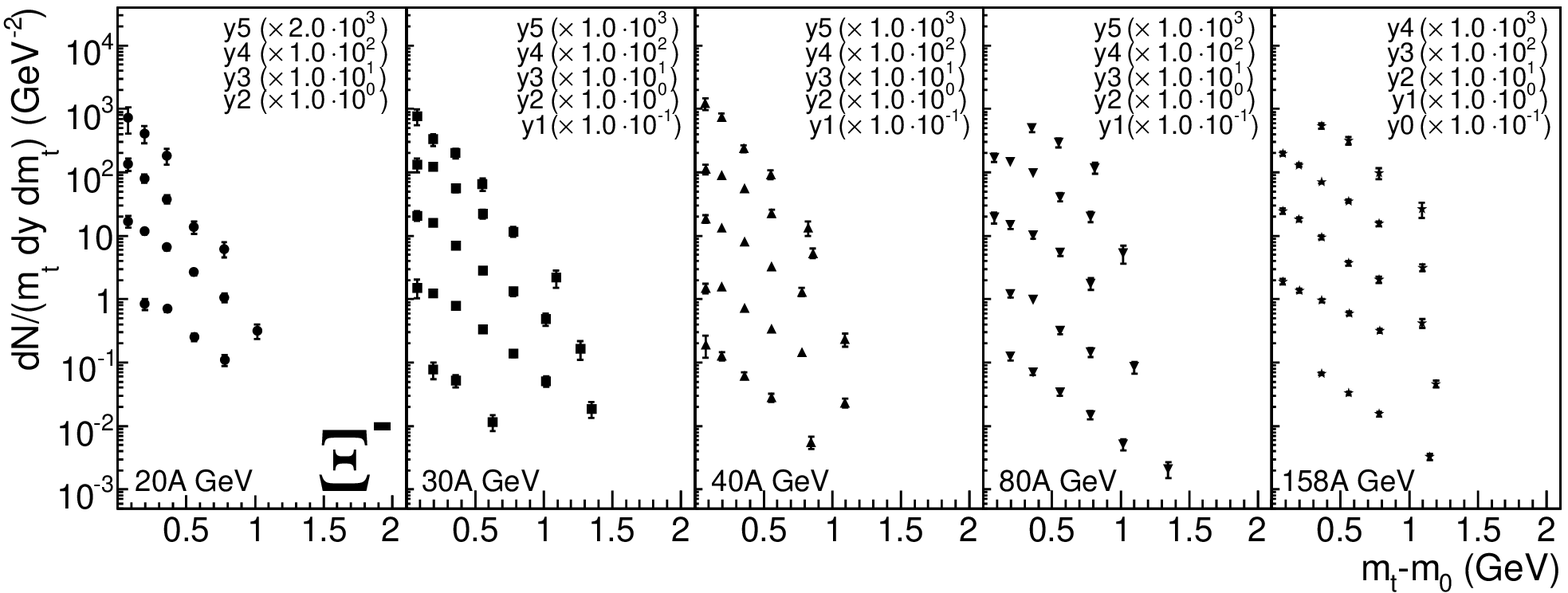}
\end{minipage}
\vspace*{-3.7mm}
\vspace*{-4.8mm}
\begin{minipage}[b]{0.95\linewidth}
\includegraphics[width=0.95\linewidth]{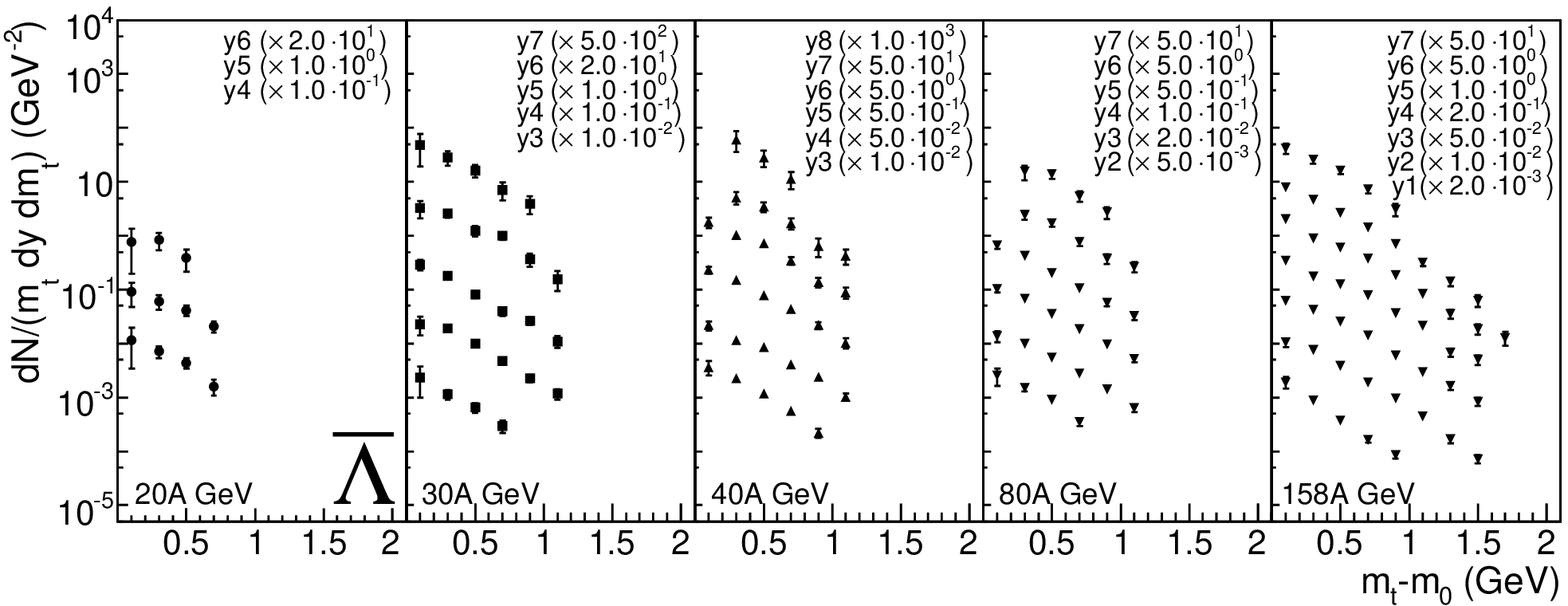}
\end{minipage}
\vspace*{-3.7mm}
\vspace*{-4.8mm}
\begin{minipage}[b]{0.95\linewidth}
\includegraphics[width=0.95\linewidth]{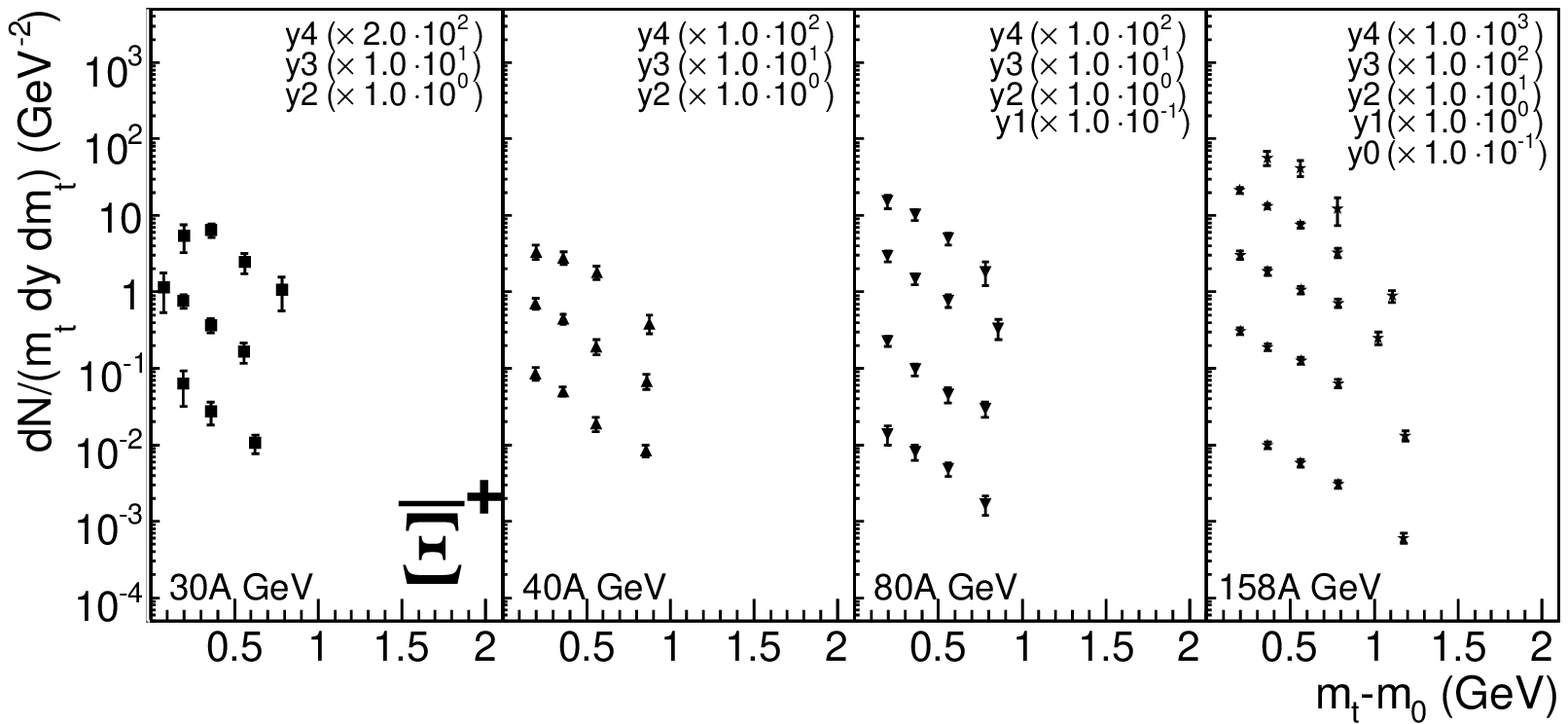}
\end{minipage}
\caption{\label{fig:all_mt} 
The transverse mass spectra of \lam, \xim, \lab, and \xip\ for central 
Pb+Pb collisions in different rapidity bins at 20$A$, 30$A$, 40$A$, 
80$A$, and 158\agev.  Every rapidity bin has a width of 0.4.  Bin 
$y0$ starts at -2.0.  The data points are scaled for clarity.  Only 
statistical errors are shown.}
\end{figure*}
%

The transverse mass spectra of \lam\ and \lab\ measured around mid-rapidity
($|y| < 0.4$) are shown in \Fi{fig:lm_lb_midrap_mt} and the ones of \xim\ 
and \xip\ ($|y| < 0.5$) in \Fi{fig:xi_xb_midrap_mt}.  The \mt~spectra were 
fitted by an exponential in the transverse mass range $\mtmzero >  0.2$~\gevcc\
as defined in \Eq{eq:expo}.  The resulting inverse slope parameters $T$ are 
summarized in \Ta{tab:summary}.  Due to the significant curvature of the 
\mt~spectra equation \Eq{eq:expo} does not provide a satisfactory description 
of the data over the whole \mt~range.  Therefore, the \mt~spectra were 
additionally fitted by a blast wave model which assumes a transversely 
expanding emission source \cite{BLASTWAVE}.  The parameters of this model 
are the freeze-out temperature \tf\ and the transverse flow velocity \betas\ 
at the surface.  Assuming a linear radial velocity profile 
$\betat (r) = \betas\; r/R$, which is motivated by hydrodynamical calculations, 
the \mt~spectrum can be computed from
\begin{equation}
  \label{eq:blast}
  \!\! 
  \frac{1}{\mt} \frac{\der N}{\der \mt \der y} 
  \propto 
  \! \int_{0}^{R} \!\!\! r \: \der r\; \mt \:
  \textrm{I}_{0} \!\!\left(\!\frac{\pt \sinh \rho}{\tf} \!\right)\!
  \textrm{K}_{1} \!\!\left(\!\frac{\mt \cosh \rho}{\tf} \!\right),
\end{equation}
where $R$ is the radius of the source and $\rho = \tanh^{-1}\! \betat$
is the boost angle.  
Since the measurements for the different particle species do not provide 
an equally good constraint on the fit procedure if both parameters are
allowed to vary freely, the transverse flow velocity was fixed to 
$\btavg = 2/3 \: \betas = 0.4$.  The results of the fits are shown as 
dotted lines in \Fis{fig:lm_lb_midrap_mt}{fig:xi_xb_midrap_mt} and the 
obtained fit parameters \tf\ are listed in \Ta{tab:blastwave}.  They
turn out to be significantly lower for \xim\ than for \lam\ at all
beam energies.  This difference is also visible for the anti-particles,
although less pronounced.  Even though this observation is based on
a relatively simple model, it might indicate that the transverse mass
spectra of \lam\ and $\Xi$ are not determined by the same kinetic 
freeze-out condition.

%
\begin{table}[ht]
\caption{\label{tab:blastwave}
The parameter \tf\ resulting from the fit with the blast wave model.
\btavg\ was fixed in all cases to 0.4.  \tf\ is given in MeV.  Errors
are statistical only.}
\begin{ruledtabular}
\begin{tabular}{lcccc}
Beam energy     & 
  \tf(\lam)     & 
  \tf(\lab)     & 
  \tf(\xim)     & 
  \tf(\xip)     \\ 
                \hline
20\agev\  & 100$\pm$2  & 166$\pm$38 &  82$\pm$7 & ---        \\
30\agev\  & 107$\pm$1  & 134$\pm$9  &  83$\pm$5 & 122$\pm$30 \\
40\agev\  & 115$\pm$2  & 143$\pm$7  &  82$\pm$4 & 127$\pm$17 \\
80\agev\  & 121$\pm$2  & 136$\pm$6  &  95$\pm$8 & 108$\pm$12 \\
158\agev\ & 140$\pm$2  & 146$\pm$3  & 109$\pm$5 & 156$\pm$9  \\
\end{tabular}
\end{ruledtabular}
\end{table}
%

To allow for a model independent study of the energy dependence of \mt~spectra, 
the averaged transverse mass \mtavg\ was calculated.  Since for \lam, \lab, 
and \xim\ essentially the whole range down to \mtmzero~=~0 is covered, \mtavg\ 
can be extracted from the data alone.  However, in order to extrapolate up to a 
common upper limit in \mtmzero, fit functions were used as well.  For this purpose 
two different fits were used: The blast wave model, as shown in 
\Fi{fig:lm_lb_midrap_mt}, and a fit with a double exponential (not shown) that 
also provides a good description of the data.  The different approaches allow to 
estimate the systematic error.  For \xip\ also an extrapolation to \mtmzero~=~0 
is needed.  
%
\begin{figure*}[htb]
\includegraphics[width=0.95\linewidth]{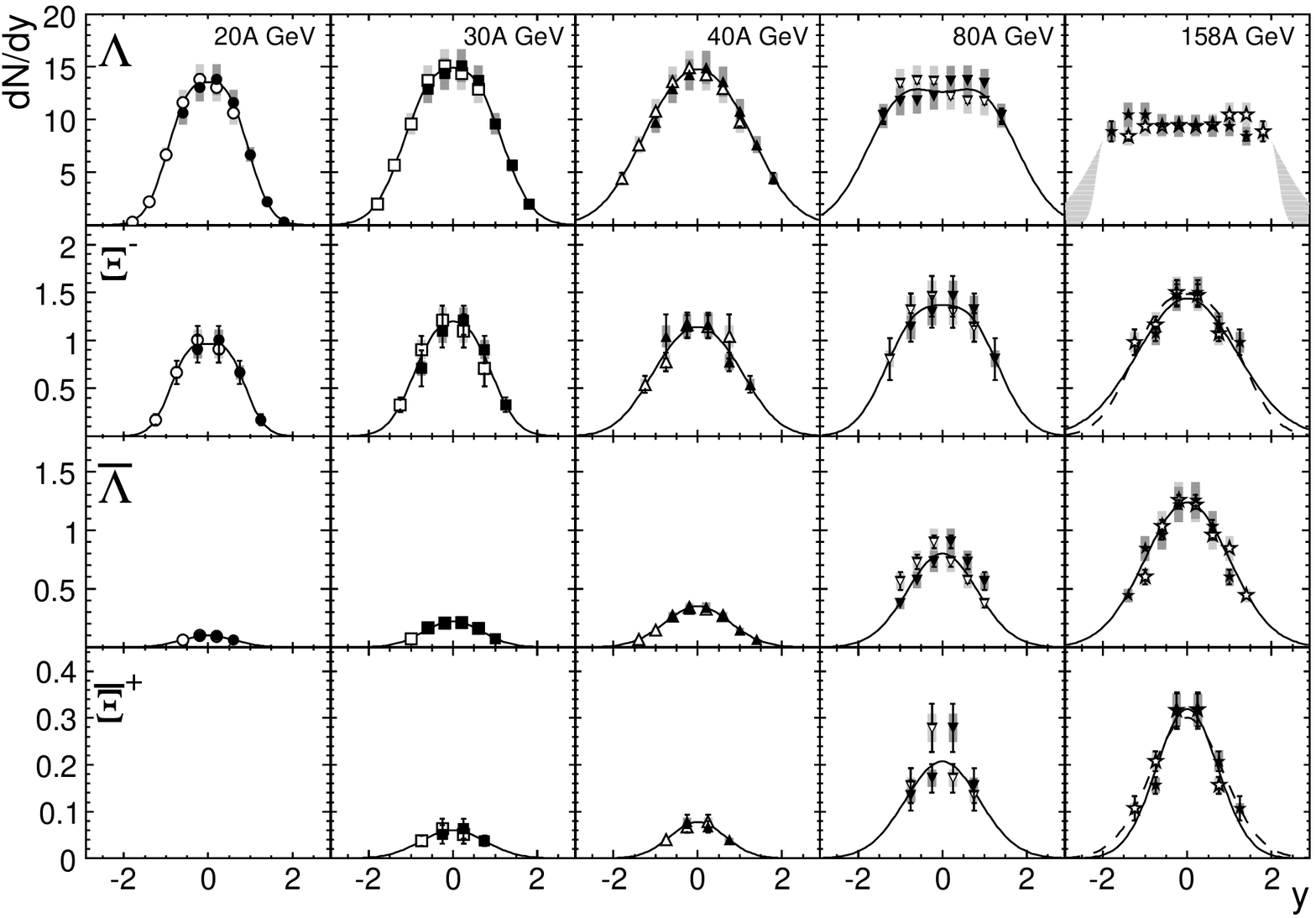}
\caption{\label{fig:all_y} 
The rapidity spectra of \lam, \xim, \lab, and \xip\ for 5 different beam 
energies. The open symbols show data points reflected around mid-rapidity. 
The systematic errors are represented by the gray boxes.  Solid lines 
are fits to the data points, used to extract the total yields, while 
dashed lines are the fits to \xim\ and \xip\ from \cite{NA49XI}.  The 
gray area in the \lam\ spectrum at 158\agev\ depicts the uncertainty 
due to the different extrapolations (see text).}
\end{figure*}
%

\clearpage

The resulting values for \mtavg, corresponding to an interval 
$0 \le \mtmzero \le 2$~\gevcc, are listed in \Ta{tab:summary}.  

Figure~\ref{fig:meanmt} shows the dependence of \mtavg\ on \sqrts\ for the
data presented here in comparison to measurements done by NA57 at the 
SPS~\cite{NA57HY40,NA57HY158}, by E896 and E917 at the AGS~\cite{E896LAM,E917LAB} 
and by STAR and PHENIX at RHIC~\cite{STARLM130,STARXI130,PHNXLM130}.  The \mtavg\
values derived from the NA57 spectra agree with the NA49 results.  In the 
SPS energy range only very little variation of \mtavg\ with \sqrts\ is observed,
followed by a slight increase towards RHIC energies.  The \mtavg\ of \lam\ is
generally higher by $\approx$~200~\mevcc\ than the one observed in p+p reactions
\cite{PPLAM} at all center-of-mass energies.  For pions, kaons and
protons a sudden change in the energy dependence of \mtavg\ around 
\sqrts~=~7~--~8~GeV was observed \cite{NA49ONSET}.  Since currently no data at 
lower energies are available, it cannot be established whether a similar feature 
is present in the energy dependence of \mtavg\ for hyperons.  However, the 
remarkably small energy variation shown in \Fi{fig:meanmt} would still be in 
line with the behavior observed for the other particle species.

%
\begin{table}[ht]
\caption{\label{tab:twogaus} The parameter $\sigma$ and $y_{0}$ resulting
from the fits with the sum of two Gauss functions (see \Eq{eq:twogaus}) to
the rapidity distributions of \lam\ and \xim.}
\begin{ruledtabular}
\begin{tabular}{lcccc}
Beam energy      & 
  $\sigma$(\lam) & 
  $y_{0}$(\lam)  & 
  $\sigma$(\xim) & 
  $y_{0}$(\xim)  \\ 
                 \hline
20\agev\  & 0.51$\pm$0.01 & 0.49$\pm$0.01 & 0.45$\pm$0.08 & 0.45$\pm$0.07 \\
30\agev\  & 0.66$\pm$0.02 & 0.59$\pm$0.01 & 0.56$\pm$0.15 & 0.47$\pm$0.11 \\
40\agev\  & 0.91$\pm$0.06 & 0.65$\pm$0.04 & 0.76$\pm$0.16 & 0.54$\pm$0.12 \\
80\agev\  & 0.87$\pm$0.07 & 0.94$\pm$0.06 & 0.71$\pm$0.32 & 0.68$\pm$0.13 \\
158\agev\ & ---           & ---           & 1.18$\pm$0.18 & ---           \\
\end{tabular}
\end{ruledtabular}
\end{table}
%

The measurements on \mtavg\ are also compared to the string hadronic model
UrQMD1.3.  While this model in principle reproduces the observed near 
independence of \mtavg\ on \sqrts\ in the SPS energy region, it fails to match 
its magnitude.  The calculation is always $\approx\:$100~MeV below the data.
Additionally, this version of UrQMD does not describe the slow increase towards 
RHIC.  


\section{\label{sec:rap_spectra} Rapidity spectra}

Figure~\ref{fig:all_mt} summarizes the \mt~spectra of \lam, \xim,
\lab, and \xip\ as measured in different rapidity bins.  The data points 
cover a large fraction of the phase space and thus allow to extract 
rapidity distributions by integrating the transverse mass spectra.  
Table~\ref{tab:summary} summarizes the resulting rapidity densities around 
mid-rapidity and \Fi{fig:all_y} shows the resulting $y$~spectra.  For \lam\ 
a clear evolution of the spectral shape with beam energy is observed.  
While the rapidity spectrum at 20\agev\ has an almost Gaussian shape, a 
plateau around mid-rapidity is developing that widens with increasing energy.  
At 158\agev\ the spectrum is finally constant in the measured rapidity range.  
This reflects the continuous change of the rapidity distribution of the
net-baryon number in this energy range \cite{NA49QM06}.  While at lower 
energies the final state distribution of the incoming nucleons looks 
thermal, the rapidity distribution of the net-baryons develops a distinct 
minimum at mid-rapidity with increasing energy due to incomplete stopping.  
Since \lam\ carry a significant fraction of the net-baryon number they 
follow this change to a large extent.  A similar behavior, although less 
pronounced, is visible for the \xim\ as well.  \lab\ and \xip, on the other 
hand, are well described by Gaussians at all beam energies.

%
\begin{figure}[ht]
\includegraphics[width=0.8\linewidth]{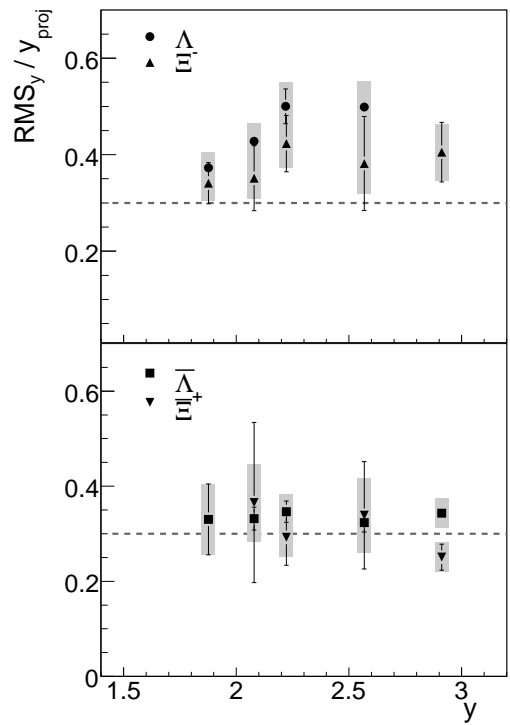}
\caption{The RMS widths of the rapidity distributions $RMS_{\rb{y}}$, 
normalized by the projectile rapidity \yproj, as a function of \yproj.  
The systematic errors are represented by the gray boxes.}
\label{fig:rmsrap} 
\end{figure} 
%

In order to determine total multiplicities, extrapolations into the 
unmeasured $y$~regions are needed.  Therefore, \lam\ and \xim\ were fitted
with a sum of two Gauss functions placed symmetrically around mid-rapidity
\begin{equation}
  \label{eq:twogaus}
  \!\! 
  \frac{\der N}{\der y} 
  \propto 
  \exp \left\{ - \frac{(y - y_{0})^{2}}{2 \sigma^{2}} \right\} +
  \exp \left\{ - \frac{(y + y_{0})^{2}}{2 \sigma^{2}} \right\} .
\end{equation}
The resulting fit parameters are listed in \Ta{tab:twogaus}.  In case of 
the \xim\ at 158\agev\ a single Gaussian turned out to provide a better 
fit to the data (solid line in \Fi{fig:all_y}).  For the \lam~distribution 
at 158\agev\ a fit cannot be performed since the measurement does not allow 
to determine the end of the \dndy~distribution. 
Here, the extrapolation has to be based on different assumptions on 
the spectral shape.  An upper limit on the contribution from the unmeasured 
parts can be derived by using the measured net-proton distribution at 158\agev\ 
\cite{NA49STOP} to describe the tails.  Another approach is to assume the 
same shape for the \lam\ rapidity distribution as has been measured for 
central S+S reactions at 200\agev\ \cite{NA35STR}, which then results in a 
lower total yield.  The multiplicity quoted in \Ta{tab:summary} is the average
between both extrapolations and their difference is taken as its systematic error.

%
\begin{figure}[ht]
\includegraphics[width=0.95\linewidth]{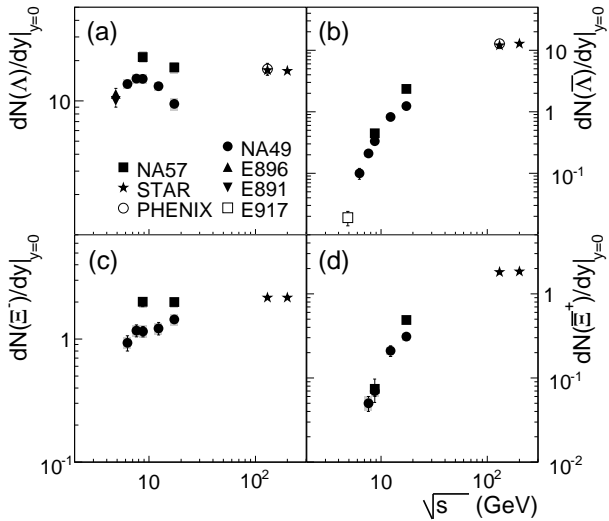}
\caption{The rapidity densities \dndy\ at mid-rapidity (\lam/\lab: 
$|y| < 0.4$, \xim/\xip: $|y| < 0.5$) in central Pb+Pb and Au+Au collisions
as a function of \sqrts.  The systematic errors are represented by gray 
areas, mostly hidden by the symbols.  Also shown are data from the NA57 
collaboration \cite{NA57HY40,NA57HY158}, as well as from AGS 
\cite{E891LAM,E896LAM,E917LAB} and RHIC experiments 
\cite{STARLM130,STARXI130,STARHY200,PHNXLM130}. }
\label{fig:dndy} 
\end{figure} 
%
  
For \lab\ and \xip\ a single Gauss function was used to derive the 
total yields.  The resulting fit parameters $\sigma$ are identical to the 
values for $RMS_{\rb{y}}$ tabulated in \Ta{tab:summary}.  The rapidity spectra 
of \xim\ and \xip\ at 158\agev\ also agree well with a fit to the previously 
published data (dashed lines in \Fi{fig:all_y}).  Figure~\ref{fig:rmsrap} 
summarizes the energy dependence of the $RMS_{\rb{y}}$ values.  While the 
widths of the \lab\ and \xip\ distributions agree with each other and exhibit 
an approximately linear dependence on the projectile rapidity \yproj\ 
($RMS_{\rb{y}} \approx 0.3 \:\yproj$, see dashed line in \Fi{fig:rmsrap}), 
the \lam\ and \xim\ show a different behavior.  Here $RMS_{\rb{y}}/\yproj$ 
is larger and also clearly energy dependent. The effect is more pronounced 
for the \lam\ than for the \xim.


\section{Particle yields}


%
\begin{figure}[ht]
\includegraphics[width=0.95\linewidth]{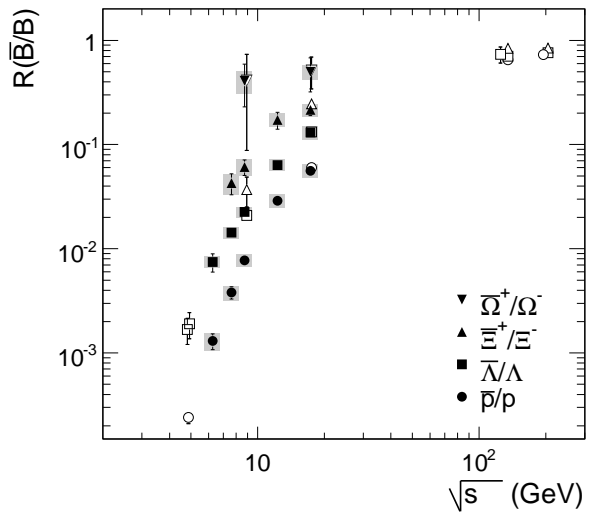}
\caption{The \pbar/p \cite{NA49PPBAR}, \lab/\lam, \xip/\xim, and 
\omplus/\ommin\ \cite{NA49OM} ratios around mid-rapidity (\lam: $|y| < 0.4$, 
$\Xi$ and $\Omega$: $|y| < 0.5$) in central Pb+Pb and Au+Au collisions as a 
function of \sqrts.  Also shown as open symbols are data from the SPS 
experiments NA44 \cite{NA44PPBAR} and NA57 \cite{NA57HY40,NA57HY158}, from 
AGS \cite{E802PPBAR,E891LAM,E896LAM,E917LAB} and RHIC experiments
\cite{STARPR130,STARLM130,STARXI130,STARPR200,STARHY200,PHNXLM130,PHNXPR200}.
The symbols are slightly displaced for clarity.}
\label{fig:bbarbratios} 
\end{figure} 
%

Figure~\ref{fig:dndy} shows the rapidity densities around mid-rapidity
as a function of \sqrts.  The energy dependence of \dndy\ for \lam\ 
exhibits a complicated structure.  It rises from AGS to a maximum at a 
beam energy of 30\agev, then drops towards top SPS energy and rises
again slowly to \sqrts~=~130~GeV.  This can be understood by an interplay
of the slow rise of the \lam\ multiplicity from \ebeam~=~30\agev\ on 
(see \Fi{fig:mult}a) and the pronounced change of shape seen in the
rapidity distribution in the same energy region (see \Fi{fig:all_y}).  
Since the \lam~yield gets distributed more and more evenly along the
rapidity axis, the mid-rapidity \dndy\ is reduced above \ebeam~=~30\agev.  
At some point the redistribution along $y$ is compensated again by the 
further increase of the \lam\ multiplicity, so that the rapidity density 
\dndy\ is again higher at RHIC.  Such a significant structure in the 
energy dependence is not observed for the \xim, where the mid-rapidity 
\dndy\ increases more smoothly by a factor of $\approx\:$2 from 
\ebeam~=~20\agev\ towards RHIC.  However, also here a small structure 
in the energy dependence is visible between 20$A$ and 80\agev.  For 
\lab\ and \xip, where no change in the shape of the \dndy~spectra is 
seen, the mid-rapidity \dndy~values increase rapidly over $\approx\:$2 
orders of magnitude between \ebeam~=~20\agev\ and \sqrts~=~130~GeV.

It should be noted that at this point there is a significant disagreement 
between the measurements presented here and the data published by the 
NA57 collaboration \cite{NA57HY40,NA57HY158}.  Even though the NA57 data 
follow the same trend in the energy dependence, they are systematically 
higher than the NA49 results \footnote{The NA57 yields have been scaled 
by the corresponding number of wounded nucleons to correct for the slightly 
different centrality selection compared to NA49.}.  This discrepancy is 
generally of the order of 1~--~2.5 standard deviations with the only 
exception of the \xip~measurements at 40\agev.  The measured particle 
ratios, on the other hand, show a good agreement between the two experiments.  
Despite intensive discussions between both collaborations, the origin of 
the discrepancies is not yet found.

%
\begin{figure}[ht]
\includegraphics[width=0.95\linewidth]{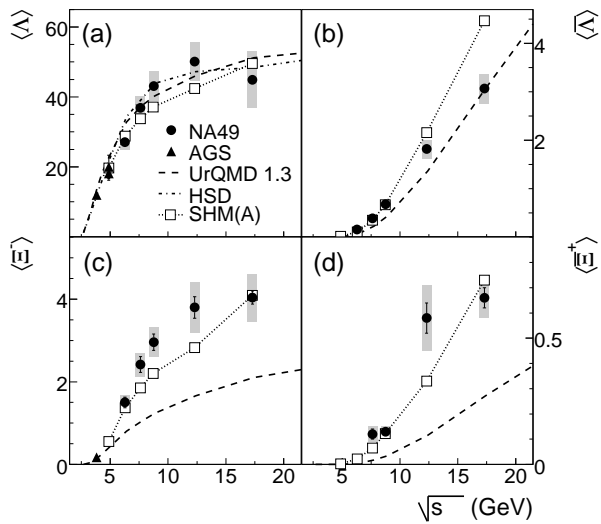}
\caption{The total multiplicities of \lam\ (a), \lab\ (b), \xim\ (c), 
and \xip\ (d) in central Pb+Pb and Au+Au collisions as a function of 
\sqrts.  The systematic errors are represented by the gray 
boxes.  Also shown are AGS data \cite{E891LAM,E896LAM,E895HYP}, as well 
as calculations with string hadronic models (HSD, UrQMD1.3 
\cite{HSD,URQMD,HSDURQMD}) and with a statistical hadron gas model 
(SHM(A)~\cite{BECATTINI4}).}
\label{fig:mult} 
\end{figure} 
%

The energy dependence of the antibaryon/baryon ratios 
$R(\bar{\textrm{B}}/\textrm{B})$, measured at mid-rapidity, are compared 
for protons, $\Lambda$, $\Xi$, and $\Omega$ in \Fi{fig:bbarbratios}.  The 
ratios exhibit a rapid rise for all particle species over several orders 
of magnitude in the SPS energy range and converge towards values close to 
1 at RHIC energies.  There is a distinct hierarchy of the ratios, depending 
on the strangeness content of the baryons: 
\begin{displaymath}
R(\omplus/\ommin) > R(\xip/\xim) > R(\lab/\lam) > R(\pbar/\textrm{p}).
\end{displaymath}
Also, the energy dependence in the SPS region gets slightly weaker with 
increasing strangeness.  The $\bar{\textrm{B}}/\textrm{B}$~ratios at
mid-rapidity directly reflect the drastic change in the net-baryon number.
However, the sensitivity depends to some extent on the valence quark content
of the baryon which is thus responsible for the observed hierarchy.

 
The total multiplicities, as determined from the \dndy~spectra shown in
\Fi{fig:all_y}, are compiled in \Fi{fig:mult} together with AGS data where
available \cite{E891LAM,E896LAM,E895HYP}.  The total multiplicities of 
\lam\ and \xim\ increase quite rapidly at lower energies, while from 
$\sqrts \approx\:$8~GeV on they rise only moderately with energy.  \lab\
and \xip, on the other hand, exhibit a continuous fast increase with beam
energy.  The measurements are confronted with several hadronic models.
In \Fi{fig:mult}a calculations with the string hadronic models HSD~\cite{HSD}
and UrQMD1.3~\cite{URQMD} for \lamavg\ as a function of \sqrts\ are shown,
as well as results from a fit with a statistical hadron gas model~\cite{BECATTINI4} 
(SHM(A)).  All three models are able to describe the data satisfactorily.  
A similar picture is observed for \labavg\ (see \Fi{fig:mult}b), although 
the fit with the statistical hadron gas model seems to overpredict the 
measurements at \ebeam~=~80$A$ and 158\agev.  The difference between 
UrQMD1.3 and the statistical model is more pronounced for \xim\ and \xip\ 
(see \Fi{fig:mult}c and d).  While the data points at SPS energies are 
above the UrQMD1.3 calculation by a factor of $\approx\:$2, the statistical 
model fit provides a qualitative description of the measurement, although 
the agreement is not perfect.

%
\begin{figure}[ht]
\includegraphics[width=0.95\linewidth]{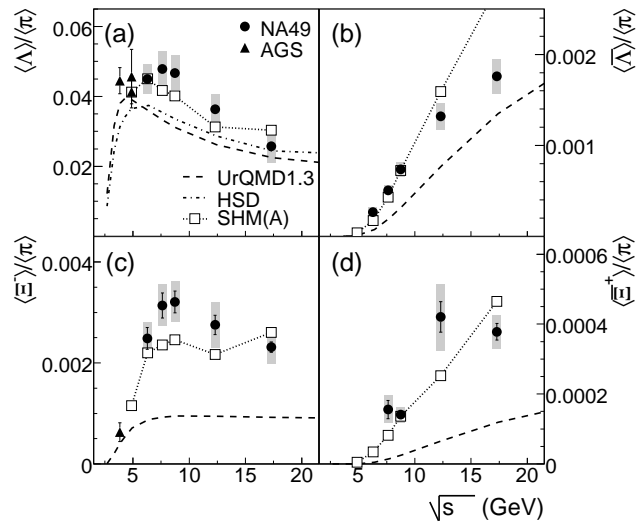}
\caption{The total multiplicities of \lam\ (a), \lab\ (b), \xim\ (c), and 
\xip\ (d) divided by the total pion multiplicities 
($\langle \pi \rangle = 1.5 \:(\langle \pi^{+} \rangle + \langle \pi^{-} \rangle)$)
in central Pb+Pb and Au+Au collisions as a function of \sqrts.  The systematic 
errors are represented by the gray boxes.  Also shown are AGS data  
\cite{E891LAM,E896LAM,E895HYP,E895PION,E802PION}, as well as calculations 
with string hadronic models (HSD, UrQMD1.3 \cite{HSD,URQMD,HSDURQMD}) and 
a statistical hadron gas model (SHM(A)~\cite{BECATTINI4}).}
\label{fig:multpion} 
\end{figure} 
%

%
\begin{figure}[ht]
\includegraphics[width=0.95\linewidth]{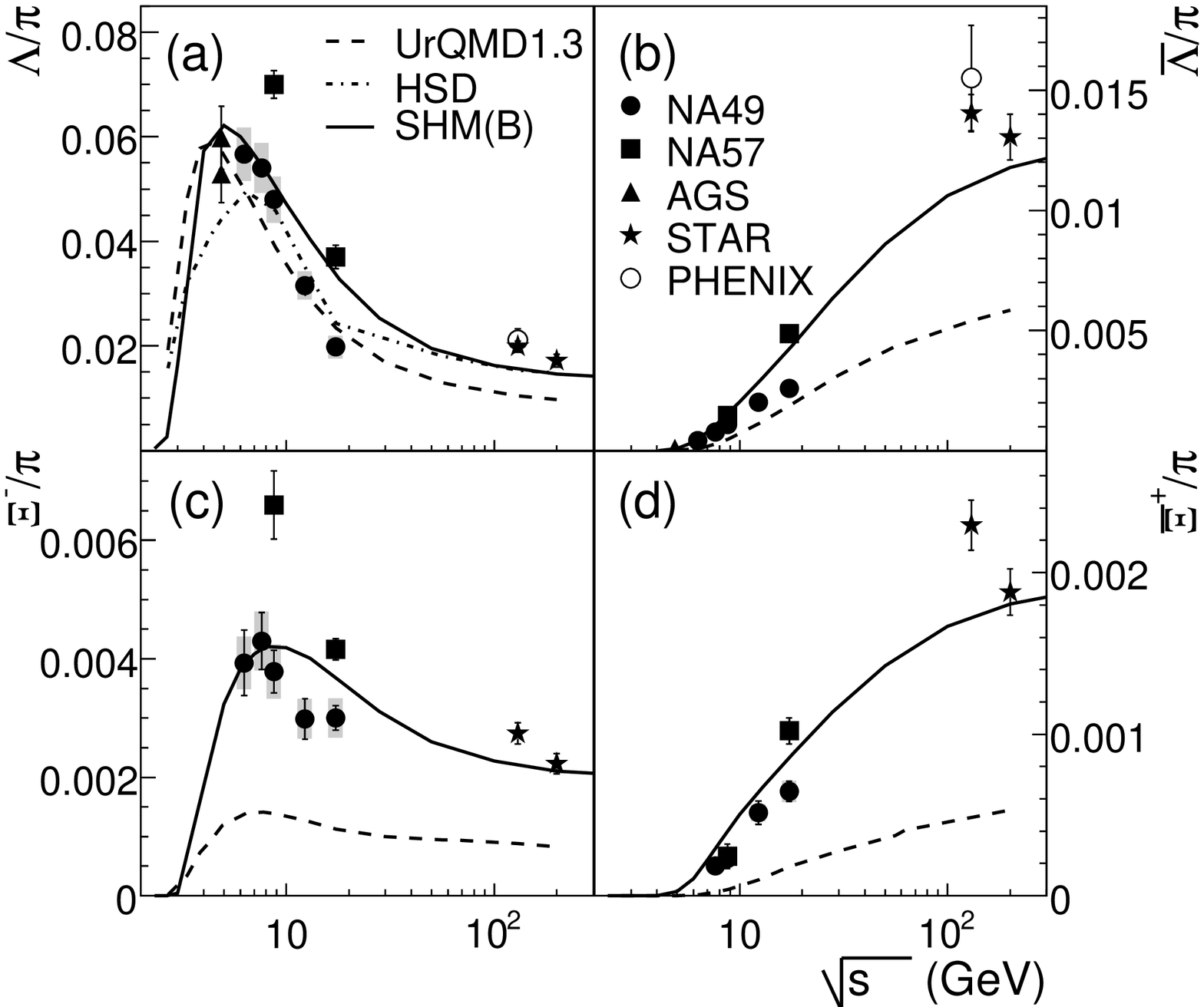}
\caption{The rapidity densities \dndy\ at mid-rapidity of \lam\ (a), \lab\ (b), 
\xim\ (c), and \xip\ (d) divided by the pion rapidity densities
($\pi = 1.5 \: (\pi^{+} + \pi^{-})$) in central Pb+Pb and Au+Au collisions as 
a function of \sqrts.  The systematic errors are represented by the gray boxes.  
Also shown are NA57 \cite{NA57HY40,NA57HY158}, AGS \cite{E891LAM,E896LAM,E917LAB,E802PION}, 
and RHIC \cite{STARLM130,STARXI130,STARHY200,STARPI130,STARPI200,PHNXLM130,PHNXPI130} 
data, as well as calculations with string hadronic models (HSD, UrQMD1.3 
\cite{HSD,URQMD,HSDURQMD}) and a statistical hadron gas model 
(SHM(B)~\cite{ANTON}).}
\label{fig:dndypion} 
\end{figure} 
%

%
\begin{figure}[ht]
\includegraphics[width=0.95\linewidth]{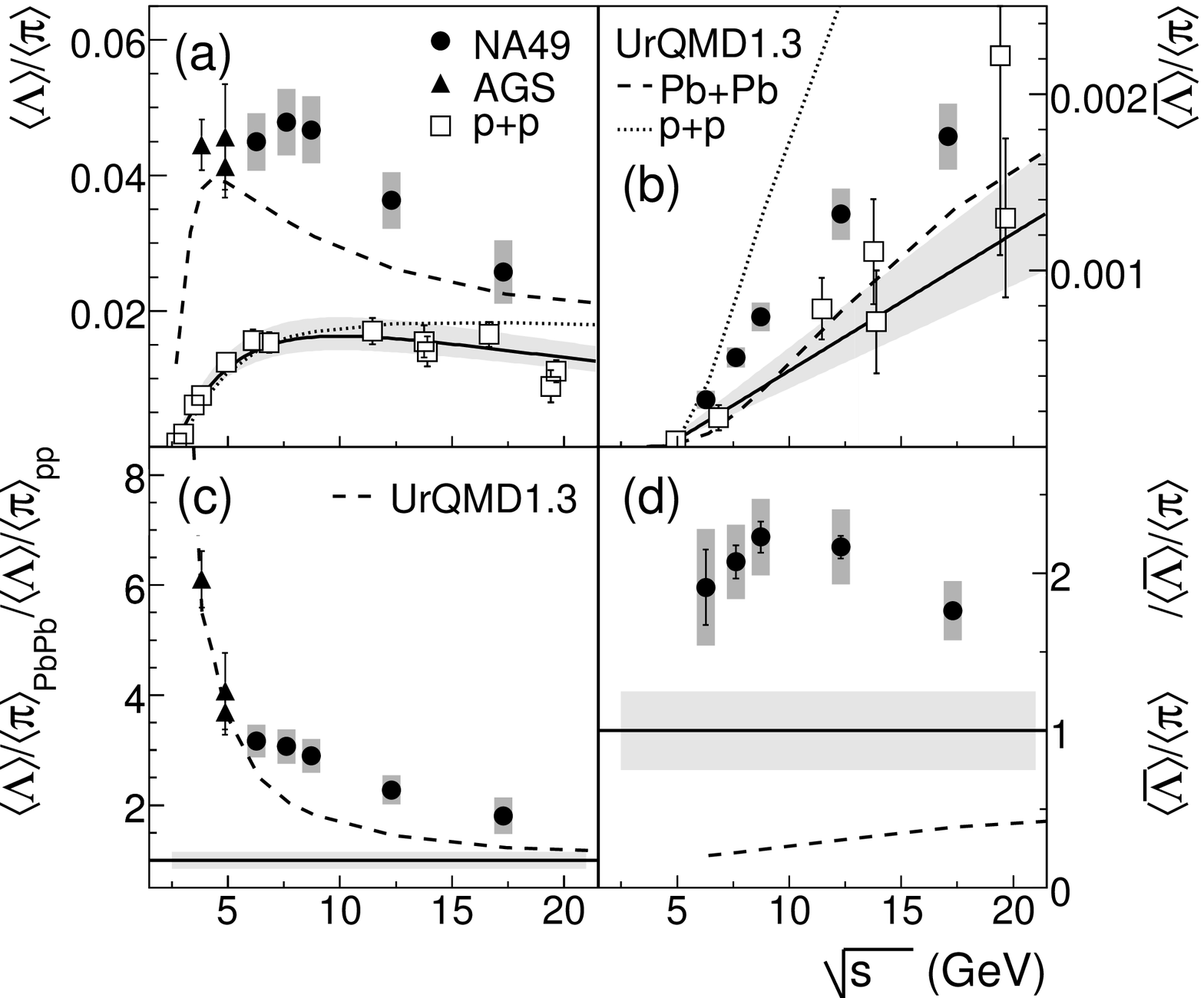}
\caption{The total multiplicities of \lam\ (a) and \lab\ (b) 
divided by the total pion multiplicities 
($\langle \pi \rangle = 1.5 \:(\langle \pi^{+} \rangle + \langle \pi^{-} \rangle)$)
for central Pb+Pb and Au+Au collisions as a function of \sqrts.  The 
systematic errors are represented by the gray boxes.  Also shown are 
AGS data \cite{E891LAM,E896LAM,E895HYP,E895PION,E802PION}, measurements 
for p+p collisions by other experiments (open squares) 
\cite{MAREKPP1,MAREKPP2}, as well as a calulation with the UrQMD1.3 model
(dashed line: Pb+Pb, dotted line p+p).  The solid line represents a 
parametrization of the p+p data (see text).  
The enhancements relative to the p+p parametrization are shown in 
panels (c) and (d).  The gray boxes denote the uncertainty of the p+p 
reference parametrization.  The dashed line represents the values from
the UrQMD1.3 model.}
\label{fig:enhance} 
\end{figure}
%

In \Fi{fig:multpion} the total multiplicities of hyperons divided by the 
total number of pions 
$\langle \pi \rangle = 1.5 \:(\langle \pi^{+} \rangle + \langle \pi^{-} \rangle)$
are compiled.  The \lamavg/\piavg\ and the \ximavg/\piavg~ratios have
distinct maxima in the region \sqrts~=~7~--~9~GeV, while the \labavg/\piavg\
and \xipavg/\piavg~ratios increase monotonously with energy.  The comparison
to the string hadronic model results of HSD and of UrQMD1.3 reveals a 
significant disagreement with the \lamavg/\piavg\ and \labavg/\piavg~ratios 
(see \Fi{fig:multpion}a and b), which is not present in the \lam\ and 
\lab~multiplicities alone as shown in \Fi{fig:mult}a and b.  This is a 
reflection of the fact that these models overpredict the pion production 
at top AGS and lower SPS energies \cite{HSDURQMD,WEBER}.  Hence, the 
disagreement with the \ximavg/\piavg\ and \xipavg/\piavg~ratios 
(\Fi{fig:multpion}c and d) is even more pronounced than for the $\Xi$ 
multiplicities alone.  The statistical hadron gas model approach provides 
overall a better description of the measured particle ratios than UrQMD1.3.  
However, the \labavg/\piavg~ratio is clearly overestimated at higher energies 
by SHM(A), while the fit results from this model are slightly below the data 
points for \ximavg/\piavg\ and \xipavg/\piavg\ for $\sqrts < 17.3$~GeV.  
In \cite{CLEYMANS1} it was argued that a statistical model approach predicts 
different positions of the maxima in the energy dependence of \lamavg/\piavg\ 
(\sqrts(max)~=~5.1~GeV) and of \ximavg/$\langle\pi^{-}\rangle$ 
(\sqrts(max)~=~10.2~GeV).  However, the existing measurements do not allow to 
determine the exact positions of the maxima with the required precision in 
order to establish a significant difference.  For this purpose also more data 
at lower energies ($\sqrts < 6$~GeV) with high precision would be required.  

Qualitatively the same picture emerges when the ratios of the mid-rapidity
yields are studied instead of the ratios of total yields, as shown in 
\Fi{fig:dndypion} together with results from RHIC experiments.  Again, the 
string hadronic models HSD and UrQMD1.3 fail to match the \xim/$\pi$ and 
\xip/$\pi$ ratios, even though a reasonable description of the \lam/$\pi$ and 
\lab/$\pi$ ratios at SPS energies is achieved, and statistical models provide 
generally a better description.  As an alternative implementation of the 
statistical hadron gas model here the one by \cite{PBM1,ANTON} (SHM(B)) is 
used.  While in SHM(A)~\cite{BECATTINI4} a separate fit at each energy to all 
available particle multiplicities is performed by varying chemical freeze-out 
temperature \tch\ and baryonic chemical potential \mub, the input parameters 
\tch\ and \mub\ in SHM(B)~\cite{ANTON} are taken from a smooth parametrization
of the \sqrts-dependence of the original fit results.  In addition, the model 
SHM(A) includes a parameter to allow for strangeness undersaturation \gams, 
which is not present in model SHM(B) (i.e. \gams~=~1).  Both models use a 
grand canonical ensemble for the results shown here.  In the case of model 
SHM(B) an additional correction by a canonical suppression factor is applied.  
However, for central A+A collisions this correction is only effective at AGS 
energies ($\sqrts \le 5$~GeV) \cite{ANTON}.  Thus, SHM(B) provides a baseline 
defining the state of maximal chemical equilibrium that is attainable.  However, 
the parametrization that provides the basis of SHM(B) has been tuned to fit 
mid-rapidity ratios, while the fits with SHM(A) have been done for total 
multiplicities which complicates a direct comparison between the two approaches.  
SHM(B), as shown in \Fi{fig:dndypion}, generally overpredicts all measured 
mid-rapidity ratios at the higher SPS energies ($\sqrts = 12 - 17$~GeV), while 
at lower SPS and at RHIC energies a satisfactory agreement is achieved.  Therefore, 
in the data a sharper maximum in the energy dependence of the \lam/$\pi$ and 
\xim/$\pi$~ratios is observed than in the model. The NA57 results\footnote{The 
NA57 yields are normalized to the corresponding NA49 pion measurements.} exibit 
a similarly shaped energy dependence. However, the ratios are generally higher 
than the NA49 results.

The observed maxima in the \lam/$\pi$ and \xim/$\pi$ ratios occur in the
same energy range as the observed distinct peak in the K$^{+}$/$\pi^{+}$ ratio 
\cite{NA49ONSET}.  Since the latter can be interpreted as a signature for the 
onset of deconfinement, the question appears whether the maxima in the \lam/$\pi$ 
and \xim/$\pi$ ratios can be attributed to the same effect.  In contrast to 
the K$^{+}$, which carry together with the K$^{0}$ the bulk of the anti-strange 
quarks and are thus a relatively direct measure of the strangeness production, 
the interpretation of the strange baryons is complicated by the fact that 
their sensitivity to the strangeness production is strongly modified by the 
energy dependent baryon number distributions.  At low energies, with high baryonic 
chemical potential, the production of baryons is favoured and more strange quarks 
will end up in \lam\ and \xim, compared to higher energies where strange quarks
might predominantly be contained in K$^{-}$ and $\bar{\textrm{K}}^{0}$. This is 
underlined by the fact that the statistical model approaches, which reflect the 
dependence of particle yields on \mub, provide a relatively good describtion of 
the data.  Whether the remaining discrepancies between SHM(B) and the mid-rapidity 
ratios at 80$A$ and 158\agev\ (see \Fi{fig:dndypion}, panels (a) and (c)) might be 
attributed to the onset of deconfinement can in the light of the systematic 
uncertainties not be definitely answered.  However, one should keep in mind that
the mid-rapidity \lam/$\pi$ and \xim/$\pi$ ratios are also strongly affected by 
the rapid change of the shape of the \lam\ and \xim\ rapidity distributions with 
energy.  This effect will cause a more pronounced energy dependence of the 
mid-rapidity ratios in comparison to the 4$\pi$ ratios, which in principle cannot 
be described by statistical models.

The \lamavg/\piavg\ and \labavg/\piavg~ratios, as measured in central 
nucleus--nucleus collisions, are compared to data obtained in p+p collisions
\cite{MAREKPP1,MAREKPP2} in \Fi{fig:enhance}.  The p+p measurements were 
parametrized by a fit function.  For \lamavg/\piavg\ the following function 
with the fit parameters $a$, $b$, and $c$ was used:
\begin{equation}
\lamavg/\piavg(p+p) = c \: [ 1 - \exp(-(\sqrt{s} - \sqrt{s_{0}}) / a) 
                              + b (\sqrt{s} - \sqrt{s_{0}}) ]
\end{equation}
Here, $\sqrt{s_{0}}$ denotes the threshold center-of-mass energy.  The result 
of the fit is displayed in \Fi{fig:enhance}a.  It provides a reasonable 
description of the available data in the energy range of $\sqrts < 20$~GeV.  
Similarly, the energy dependence of the \labavg/\piavg~ratio was parametrized 
by a straight line.  However, the existing measurements are much less precise 
than in the \lamavg/\piavg\ case.  Based on these parametrizations, the energy 
dependence of an enhancement factor $E$ relative to p+p, defined as
\begin{equation}
E = \langle N \rangle/\piavg|_{A+A} \left/ 
\langle N \rangle/\piavg|_{p+p} \right.
\end{equation}
can be determined.  As shown in \Fi{fig:enhance}c, the enhancement factor
for \lam\ exhibits a clear increase from a factor of $\approx\:$2 to $>\:$3 
towards lower energies.  For $\sqrts < 4$~GeV the AGS measurement of 
\cite{E895HYP} suggests an even more dramatic rise towards very low energies.  
For \lab\ the enhancement is of the order of $\approx\:$2, without any 
significant energy dependence in the range covered by the data.
While the UrQMD1.3 model qualitatively reproduces the energy dependence 
of the $\lam$-enhancement, it fails to describe the enhancement of \lab.
In fact, the model rather predicts a \lab-suppression, which is mainly due to
the fact that the \labavg/\piavg~ratio in p+p reactions is grossly overestimated
(see dotted line in \Fi{fig:enhance},b).  Since the net-baryon density is 
largest around \sqrts~=~5~GeV, the production of strange baryons exhibits a 
pronounced maximum at these energies.  This effect is described by all hadronic
models considered here and consequently the \lam/$\pi$-ratios are well reproduced
(\Fis{fig:multpion}{fig:dndypion}).  Moreover, the energy dependence of the
$\lam$-enhancement seems to be affected by the redistribution of the baryon number,
which is suggested by the fact that UrQMD1.3 gives a similar increase towards 
low energies.  In comparison, the doubly strange \xim\ is less sensitive to the
baryon number density and more to the overall strangeness production, which
may explain why string hadronic models fail to describe the data.  For the 
corresponding antiparticles this argument applies even more strongly.
Whether the $\Xi$-enhancement also increases towards low energies, similar to
the \lam, can currently not be decided due to the lack of precise reference data 
in p+p at lower energies.


\section{Summary}

A systematic study of the energy dependence of \lam, \lab, \xim, and \xip\
production in central Pb+Pb reactions at SPS energies is presented.  

The shape of the \mt~spectra exhibits only a weak dependence on beam energy, 
which is also reflected in the moderate increase of \mtavg\ towards the
higher RHIC energies.  A similar behavior was also observed for pions, kaons, 
and protons.  For these particles a sudden change in the energy dependence
around \sqrts~=~7~--~8~GeV was found in addition.  Due to the lack of 
data at lower energies it currently cannot be established whether a similar 
feature is present in the energy dependence of \mtavg\ for hyperons.  There 
is an indication for a slightly weaker energy dependence of \mtavg\ for 
\lab\ than for \lam, the values for \lab\ being above the ones for \lam.  
Generally, the measured \mtavg\ is higher for all investigated particle 
species than what is predicted by the UrQMD1.3 model.  

For \lam\ and \xim\ rapidity spectra a clear change of the shape is 
observed.  The almost Gaussian like spectral form develops a plateau
around mid-rapidity towards higher energies, reflecting the change of the 
longitudinal distribution of the net-baryon number.  The rapidity spectra
of \lab\ and \xip, on the other hand, can be described by single Gaussians
at all investigated energies, whose $\sigma$ increases monotonically
with energy.

Also the energy dependence of the total yields shows a distinct difference
between baryons and anti-baryons.  While for the \lab\ and \xip\ 
multiplicities a continuous rapid rise with beam energy is observed, 
the increase of the \lam\ and \xim\ yields is clearly weaker above 
\sqrts~=~7~--~8~GeV than below.  This difference gets even more pronounced 
when dividing the total multiplicities of the hyperons by those of pions.
The energy dependence of the \lamavg/\piavg\ and \ximavg/\piavg\ ratios
exhibits significant maxima in the region $5 < \sqrts < 10$~GeV, while 
the \labavg/\piavg\ and \xipavg/\piavg\ ratios increase monotonically.
The total multiplicities of \lam\ and \lab\ are well described by the
string hadronic UrQMD1.3 model.  However, \xim\ and \xip\ multiplicities
are underpredicted by factors of 2~--~3 at SPS energies.  A better overall
description of all measured yields is provided by statistical hadron gas
models.


\begin{acknowledgments}
This work was supported by the US Department of Energy
Grant DE-FG03-97ER41020/A000,
the Bundesministerium f\"{u}r Bildung und Forschung, Germany (06F137), 
the Virtual Institute VI-146 of Helmholtz Gemeinschaft, Germany,
the Polish State Committee for Scientific Research (1 P03B 006 30, 
N N202 078735, 1 PO3B 121 29, 1 P03B 127 30),
the Hungarian Scientific Research Foundation (T032648, T032293, T043514),
the Hungarian National Science Foundation, OTKA, (F034707),
the Polish-German Foundation, the Korea Research Foundation 
(KRF-2007-313-C00175) and the Bulgarian National Science Fund (Ph-09/05).
\end{acknowledgments}



\end{document}